\documentclass[11pt,preprintnumbers,amsmath,amssymb,onecolumn]{revtex4}

\usepackage[version=4]{mhchem}
\usepackage{amsmath}
\usepackage[utf8]{inputenc}

\usepackage{cleveref}
\usepackage[driverfallback=dvipdfm]{hyperref}
\usepackage{color}
\usepackage{siunitx}
\usepackage[inline]{enumitem}
\usepackage{graphicx}
\usepackage{caption}
\usepackage{subcaption}
\graphicspath{ {./home/cbb/tex files} }
\usepackage{makeidx}
\usepackage{bm}
\usepackage[title]{appendix} 
\usepackage{tabu}
\usepackage{amssymb}
\usepackage{float}

\begin{document}

\title{\large {Modelling the role of media induced fear conditioning in mitigating post-lockdown COVID-19 pandemic: perspectives on India}}
\author{Spandan Kumar$^{1\dagger}$, Bhanu Sharma$^{2\dagger}$, Vikram Singh$^{3}$}
\email{vikramsingh@cuhimachal.ac.in, $\dagger$Equal contribution}
\affiliation{$^1$ School of Social Sciences, Indira Gandhi National Open University, New Delhi, India.\\
$^2$ Department of Biophysics, University of Delhi South Campus, New Delhi, India.\\ $^{3}$ Centre for Computational Biology and Bioinformatics, Central University of Himachal Pradesh, Dharamshala, India. }

\begin{abstract}{\noindent}\textbf{Abstract}
Several countries that have been successful in constraining the severity of COVID-19 pandemic \textit{via} ``lockdown" are now considering to slowly end it, mainly because of enormous socio-economic side-effects. An abrupt ending of lockdown can increase the basic reproductive number and undo everything; therefore, carefully designed exit strategies are needed to sustain its benefits post upliftment. To study the role of fear conditioning on mitigating the spread of COVID-19 in post-lockdown phase, in this work, we propose an age and social contact structures dependent Susceptible, Feared, Exposed, Infected and Recovered (SFEIR) model. 
Simulating the SFEIR model on Indian population with fear conditioning \textit{via} mass media (like, television, community radio, internet and print media) along with positive reinforcement, it is found that increase in fraction of feared people results in significant decrease in the growth of infected population. 
The present study suggests that, during post-lockdown phase, media induced fear conditioning in conjunction with closure of schools for about one more year can serve as an important non-pharmaceutical intervention to substantially mitigate this pandemic in India. 
The proposed SFEIR model, by quantifying the influence of media in inducing fear conditioning, underlies the importance of community driven changes in country specific mitigation of COVID-19 spread in post-lockdown phase.
\end{abstract}

\maketitle



\section{Introduction}
Originating in the Wuhan city of China with its first case reported in December 2019 \cite{whocovid19}, the recently emerged pandemic COVID-19 due to novel severe acute respiratory syndrome coronavirus 2 (SARS-CoV-2) virus has claimed more than 340,000 human lives worldwide as of $25^{th}$ May, 2020 \cite{coronavirus}. By now, this pandemic has severely affected various countries across the globe forcing more than one-third of the world population into a condition of lockdown, \textit{i.e.} suppression of people gatherings in any form that includes shutting down of schools, workplaces and public transports etc.. Although, lockdown is playing a very important role in controlling the number of infections and  mitigating the epidemic, it is not an economically viable solution, and therefore can not be sustained for long. Designing of strategies about the implementable exit scenarios and developing a blueprint of the social life in post-lockdown phase, until vaccine or treatment becomes available, is the utmost need of the hour \cite{zhigljavsky2020comparison, gilbert2020preparing}.

Human behaviour is thought to play an important role in spread of a pandemic \cite{funk2009spread, del2005effects}. One of the important factors that affect behavior is fear \cite{riva2014pandemic}. Modelling studies has shown that fear has significant effect on reducing the severity of a pandemic \cite{poletti2011effect,  perra2011towards, epstein2008coupled}. Fear has been shown to directly correlate with increase in social distancing behaviour as well as taking more precautions \cite{kim2015public, cowling2010community}.  It is reported that both social interactions and anxiety levels are controlled by projections from basolateral amygdala to medial prefrontal cortex \cite{felix2016bidirectional}. Inhibition of this projection via opto-genetics is found to decrease anxiety levels and increase social interactions in mice. Thus, information based fear conditioning can serve as an important method in conjunction with other methods that can increase social distancing and precautionary behaviour. Fear conditioning, a subtype of classical conditioning \cite{pavlov2010conditioned}, is defined as a relationship developed between aversive events and an environmental stimuli by an individual \cite{maren2001neurobiology, davis1992role}. For example, in case of COVID-19, an individual can be considered fear conditioned if he establishes relationship between illness and touching of surfaces in public place. Thus, he starts considering exposure to surfaces in public place as an exposure to virus itself. For fear conditioning to happen, an unconditional stimuli (exposure to virus in the current example) is to be paired with a conditional stimuli (touching surfaces in public space). Once this learning has taken place, conditional stimuli (surfaces in public place) will also elicit the same response in an individual even in the absence of unconditional stimuli. If the individual is subjected to conditional stimuli without unconditional stimuli for long time, the learned behavior starts diminishing. Therefore, a term ``fear-time" is defined as the time it takes for behaviour due to fear conditioning to get extinct. Operant conditioning is defined as a kind of learning in which a behavior is enhanced or reduced by methods of punishment or reinforcement. Positive reinforcement happens when a person is rewarded for a certain behavior and is thereby encouraged to repeat the same behavior \cite{skinner2019behavior}. For example, in case of COVID-19, if an individual having a belief that `touching surfaces in public places' is wrong is rewarded can increase the fear-time.

In this work, we propose a model to examine the role of fear in managing the COVID-19 spread in post-lockdown phase, with media induced fear conditioning in conjunction with operant conditioning as one such prospective method. Fear, in the present work, is defined as learning of relationship between COVID-19 (aversive event) and the various ways through which it can spread (stimuli). Fear can be learned by individuals through various means, like, seeing a person who is suffering from COVID-19, but the fastest way for such conditional learning to spread across population is through mass media \cite{towers2015mass}. Media influence plays a significant role in spreading useful information through various sources, like,  television, community radio, internet and print media (such as, newspapers, magazines) etc., resulting in changed behaviour of a community \cite{wakefield2010use}, thereby influencing the progression of a pandemic \cite{collinson2014modelling, collinson2015effects}. It has been showed that if media publicity is focussed on guiding people's behaviour, it can have significant effect on spread of pandemic \cite{yan2016media}. Individuals who interact quite frequently have stronger connection to influence others, and pass on the similar information among them \cite{bakshy2012role}. Therefore, the model developed in this paper examines the media induced fear on epidemic progression by hypothesizing that inducing fear conditioning and increasing its fear time by operant conditioning via mass media (television, community radio, internet and print media) will have a negative effect on progression of COVID-19. As we are only considering positive reinforcement; three terms, operant conditioning, reward, and positive reinforcement have been used interchangeably throughout this paper.

Rest of the paper is organized as follows: in the Methods section, we detail the design principles of proposed age- and contact-structure dependent SFEIR (Susceptible, Feared, Exposed, Infected and Recovered) model and discuss about the data sources, assumptions made as well as the simulation protocol. In the Results section, simulation outcomes of the developed SFEIR model incorporating fear conditioning \textit{via} mass media in Indian population are presented and their implications are elaborated. We first compare the simulations of SFEIR model with the classical SEIR models (not having fear conditioning) with or without lockdown, and then study the effect of varying the impact of fear conditioning and reward parameters. We then study the effect of school closure and increased fear conditioning. Finally, we conclude with the discussions and potential policy perspectives of the present work on the influence of media on collective psychology of a community to counter COVID-19 progression in the post-lockdown phase. 

\section{Methods}

\begin{figure}[!htb]
\includegraphics[height=5cm, width=10.5cm]{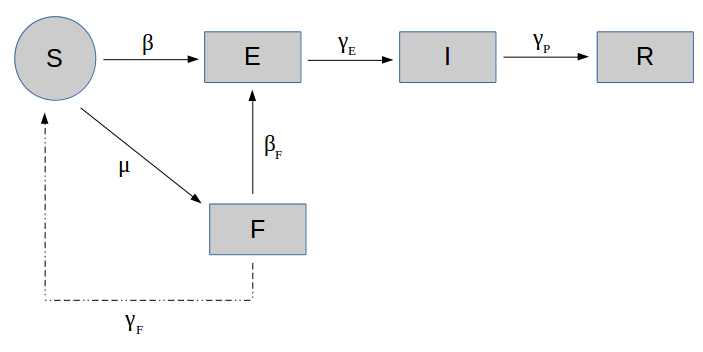}
\caption{\textbf{Flow diagram of the proposed SFEIR model}. It consists of 5 compartments \textit{viz}, Susceptible, Feared, Exposed, Infected and Recovered, and in each compartment, people belonging to all the 16 age groups may exist.}
\label{F1}
\end{figure}

\subsection{Construction of SFEIR model}
To study the effects of fear conditioning $via$ mass media in countering post-lockdown spread of COVID-19 infections, we have developed an age- and contact-structured Susceptible Feared Exposed Infected Recovered (SFEIR) model shown in Fig. \ref{F1}. A susceptible person can be exposed to infection either directly or after moving into the fear compartment \cite{epstein2008coupled}. Feared individuals are less social \cite{felix2016bidirectional} and take more precautions, and therefore will have lower probability of transmission ($\beta_{F}$) as compared to individuals in susceptible group ($\beta$). Feared individuals recover from fear with fear recovery rate $\gamma_{F}$ and move back into susceptible compartment. $\gamma_{F}$ is inversely proportional to fear time $T$. Once exposed, the rate of getting infected by pathogen is $\gamma_{E}$ and the rate of recovery is $\gamma_{P}$. \newline

While formulating the SFEIR model, age- and social contact-structures are incorporated as detailed in \cite{prem2017projecting, singh2020age}. In the age-structured population, first group consists of individuals of age $(0-4)$ years, second group is of age $(5-9)$ years, and so on, till group $16^{th}$ that consists of individuals of age $(75-79)$ years. It is considered that susceptible individuals can develop fear through various media platforms, like, television, community radio, internet, print media etc. in the post-lockdown phase. Considering $S_i$, $F_i$, $E_i$, $I_i$ and $R_i$ as the variables representing the number of individuals belonging to $i^{th}$ age group in Susceptible, Feared, Exposed, Infected and Recovered compartments, respectively, the generalized age- and social contact-structured SFEIR model is given as: 

\begin{equation}
\begin{aligned}
  \frac{dS_{i}}{dt}&= - \lambda_{i}  S_{i} -   \mu_{i} S_{i} +   \gamma _{F}(t) F_{i} \\
  \frac{dF_{i}}{dt}&=   \mu_{i}  S_{i} -  \lambda_{F_{i}} F_{i} -  \gamma _{F}(t) I_{F_{i}} \\
  \frac{dE_{i}}{dt}&=  \lambda _{i} S_{i} +  \lambda _{F_{i}} F_{i} -  \gamma _{E} E_{i} \\
  \frac{dI_{i}}{dt}&=  \gamma_{E} E_{i} - \gamma_{P}  I_{i} \\
  \frac{dR_{i}}{dt}&= \gamma_{P} I_{i} 
\end{aligned}
\end{equation}

where,  
\begin{equation*}
\begin{aligned}
\mu_{i} &=  \sum_{m=1}^M \kappa_{im}(t) *\rho_{im}(t)*\sigma_{im}(t) \\
\beta_{F}&= \beta_{Fmin} + (\beta - \beta_{Fmin})e^{-\mu_{i}.(t-t_{1})} \\
\gamma_{F}&= \frac{1}{T_{C}+[r*\tau*(1-e^\frac{-t}{\tau})]} \\ 
\end{aligned}
\end{equation*}

\begin{equation*}
\begin{aligned}
\lambda_{i}&= \beta  \sum_{j=1}^N [C_{ij}^H + C_{ij}^W+C_{ij}^S+C_{ij}^O]*(\frac{I_{j}}{N_{j}}) \\
\lambda_{Fi}&= \beta_{F}  \sum_{j=1}^N [C_{ij}^H + C_{ij}^W+C_{ij}^S+C_{ij}^O]*(\frac{I_{j}}{N_{j}}) \\
N_{i}&= S_{i} + F_{i} + E_{i} + I_{i} + R_{i} 
\end{aligned}
\end{equation*}

The rate of fear conditioning in age group $i$, $\mu_{i}$, is considered as a product of following three terms,
\begin{itemize}
\item[(i)] $ \kappa_{im}$: fraction of people who are fear conditioned by media $m$ in age group $i$ out of people who trust media $m$ in age group $i$.
\item[(ii)] $ \rho_{im}$: fraction of people who trust media $m$ in age group $i$ out of people who are exposed to media $m$ in age group $i$.
\item[(iii)] $ \sigma_{im}$: fraction of people who are exposed to media $m$ in age group $i$ per day out of total population in age group $i$.
\end{itemize}
and is given as,
\begin{equation}
\mu_{i} =  \sum_{m=1}^M \kappa_{im}(t) *\rho_{im}(t)*\sigma_{im}(t)
\end{equation}
where, $m$ represents the type of mass media (\textit{i.e.} television, community radio, internet or print media) and $M$ represents the number of mass media types. Since in this work four types of mass media are considered, so $M$=4. Under condition $\mu_{i}$ = 0, \textit{i.e.} if rate of fear conditioning is zero, for every group $i$, SFEIR will get reduced to an age- and contact-structured SEIR model which will be a variant of classical SIR model \cite{prem2017projecting, kermack1927contribution}. In this scenario, variables $\beta_{F}$ and $\gamma_{F}$ will have no effect on the resulting dynamics.

Susceptible person in group $i$ can move into exposed group after coming in contact with infected individual with probability of transmission $\beta$ and contact matrix $C_{ij}$ defined as per \cite{prem2017projecting}. Fearful individual in group $i$ will be exposed with a reduced probability of transmission $\beta_{F}$. Probability of transmission of feared individual to the exposed group is taken as time-dependent along the lines of \cite{tang2020updated, eikenberry2020mask}, as it shall decrease with the increase in the number of individuals in fear group. Hence, rate of decrease in probability of transmission ($\beta_{F}$) should be proportional to the rate of fear conditioning ($\mu_{i}$). We model the $\beta_{F}$ as,
\vspace{-4mm}
\begin{equation}
\begin{aligned}
\beta_{F} &= \beta_{Fmin} + (\beta - \beta_{Fmin})e^{-\mu_{i}.(t-t_{1})} 
\end{aligned}
\end{equation}
where,
\begin{equation*}
\begin{aligned}
\beta_{Fmin} &= \epsilon * \beta
\end{aligned}
\end{equation*}

\noindent $\epsilon$ is a constant, having value between 0 and 1, which converts the minimum probability of transmission from fear group to exposed group in terms of $\beta$. For example, if $\epsilon$ = 0.1, it means that minimum probability of transmission from fear group is 10\% to that from susceptible group.

Fear time $T$ is defined as the average time-interval an individual spends being fear conditioned \textit{i.e.} the average time from induction of fear conditioning to its extinction. Reinforcement is known to modulate conditioned behaviour \cite{skinner2019behavior}, and therefore it is considered that if fear beliefs are rewarded, fear time will increase. Fear time ($T$) is the sum of following two components:
\setlist{nolistsep}
\begin{itemize}
\item[(i)] A constant term, in the absence of any reinforcement, ($T_{C}$), and, 
\item[(ii)] A variable term, depending on the positive reinforcement through reward of beliefs, ($T_{R}$).
\end{itemize}  

\begin{equation}
T = T_{C} + T_{R}
\end{equation}

Rewards can increase the frequency of a desired behavior as well as prohibit it from getting extinct, and are quantified by a reward (utility) function \cite{schultz2006behavioral}. A reward (utility) function, $r(t)$, maps the effect of reward on the behavior of an individual in a physical world to the real numbers (${\rm I\!R}$). In this work, behavioral response is quantified using fear time ($T$), which has two parts: $T_{C}$ and $T_{R}$. Fear time of individuals will be directly proportional to the rewards associated with their beliefs e.g. a praise from social, religious or political leaders for maintaining social distancing. We define a reward function, $r(t)$, that has the capability to modulate reinforced fear time, $T_{R}$. In its general formulation, a reward function will be time-dependent function because effect of rewards are also dependent on several factors, like, when it is given, how it is given, expectation of reward, etc. \cite{schultz2006behavioral, humphreys1939effect, skinner2019behavior}. However, for simplicity, the reward function $r(t)$ is used as a time-independent constant parameter in the SFEIR model and is termed as a reward parameter $r$. As no particular method of reinforcement is used in this paper, the reward parameter $r$ is treated as an abstract quantity that can modulate $T_{R}$ in following three ways:
\begin{itemize}
\item[(i)] Reinforced fear time ($T_{R}$) increases if $r$ is present, therefore the change in $T_{R}$ is considered to be directly proportional to $r$.
\item[(ii)] Even if $r$ is present for indefinite time, $T_{R}$ cannot continuously keep on increasing and must saturate at some level. 
\item[(iii)] If reward parameter $r$ becomes zero at time $t$, $T_{R}$ should start decreasing from  time $t$ onwards with time constant $\tau$.
\end{itemize}
Therefore, $T_{R}$ is modelled using the following differential equation:
\begin{equation}
\frac{dT_{R}}{dt} =  r - \frac{T_{R}}{\tau} 
\end{equation}

\noindent Solving the above equation returns the following expression of $T_{R}$,

\vspace{2mm}
\begin{equation}
T_{R} = r * \tau * (1-e^{\frac{-t}{\tau}}) 
\end{equation}

\noindent And, the rate of recovery from fear conditioning, $\gamma_{F}$, becomes: 
\vspace{2mm}
\begin{equation}
\gamma_{F}(t) = \frac{1}{(T_{C} + T_{R}(t))} = \frac{1}{T_{C}+[r*\tau*(1-e^\frac{-t}{\tau})]}
\end{equation}

\vspace{1mm}
Partitioning the social contacts of individuals into home, workplace, school and other categories \cite{prem2017projecting}, following contact matrices are used in the SFEIR model (eq 1):\setlist{nolistsep}
\begin{itemize}
\item $C_{ij}^H$ = Contact matrices between age groups $i$ and $j$ for Home
\item $C_{ij}^W$ = Contact matrices between age groups $i$ and $j$ for Workplace
\item $C_{ij}^S$ = Contact matrices between age groups $i$ and $j$ for Schools
\item $C_{ij}^O$ = Contact matrices between age groups $i$ and $j$ for Others
\end{itemize}

During lockdown, work and schools are closed, thus, $C_{ij}^W$ = $C_{ij}^S$ is equal to 0. $C_{ij}^H$ remains the same as before lockdown. $C_{ij}^O$ is assumed to decrease as a function of time after lockdown begins. Therefore, to fit to data (number of active cases), $\lambda_{i}$ during lockdown is modelled as:
 
\begin{equation}
\lambda_{i} = \beta  \sum_{j=1}^N [C_{ij}^H + \big(\alpha_{1}-\alpha_{2} *(t-t_{o})\big) *C_{ij}^O]*(\frac{I_{j}}{N_{j}}),
\end{equation}

where, $\alpha_{2}$ is the rate of decrease of probability of transmission and $t_{o}$ is the time at which lockdown starts. $\alpha_{1}.\beta $ is probability of transmission at $ t = t_{o}$ for $C_{ij}^O$.

\begin{table}[h]
\begin{center}
\caption{Values of parameters used in SFEIR model.}
\label{Table1}
\begin{tabular}{ |p{5cm}|p{5cm}|p{5cm}|}
\hline
\textbf{Parameter} & \textbf{Value} & \textbf{References} \\
\hline
$\sigma_{im}$ for $m$= television & 0.631  & \cite{televisionmedia} \\
\hline
$\sigma_{im}$ for $m$= community radio & 0.29 & \cite{communityradiomedia} \\
\hline
$\sigma_{im}$ for $m$= internet & 0.234 & \cite{internetmedia} \\
\hline
$\sigma_{im}$ for $m$= print media & 0.148 & \cite{newspapermedia} \\
\hline
$\rho_{im}$ for $m$= television & 0.41 &  \cite{trustmedia} \\
\hline
$\rho_{im}$ for $m$= community radio & 0.41 & \cite{trustmedia} \\
\hline
$\rho_{im}$ for $m$= internet & 0.34 & \cite{trustmedia} \\
\hline
$\rho_{im}$ for $m$= print media & 0.55 & \cite{trustmedia} \\
\hline
$T_{C}$ & 2 days & Assumed \\
\hline
$\tau $ & 2 days & Assumed \\
\hline
$\beta$   & $ 0.0598 \pm 0.0011 $  & Fitted \\
\hline
$\alpha_{1}$ & $ 0.3105 \pm 0.0312 $ & Fitted \\
\hline
$\alpha_{2}$ & $ 0.0073 \pm 0.0005 $ & Fitted \\
\hline
$\gamma_{E}^{-1} $ & 5.2 days & \cite{kucharski2020early} \\
\hline
$\gamma_{P}^{-1} $ & 7 days & \cite{tang2020updated} \\
\hline
\end{tabular}
\end{center}
\end{table}

\subsection{Assumptions used for simulation of SFEIR model}

\begin{itemize}
\item[1.] $\rho_{im}$ across all age groups is taken to be equal to average for total population due to non-availabilty of data. Similarly, $\sigma_{im}$ across all age groups is also taken equal to average for total population, for same reasons. To get the effect of fear conditioning because of inter group contacts, $\kappa$ was varied to simulate the model as detailed in Results section \textit{3.5}.

\item[2.] $\kappa$ and $r$ are taken independent of time to reduce the complexity of model. 

\end{itemize}

\subsection{Simulating SFEIR model on India specific COVID-19 data}
First case of COVID-19 in India was reported on $30^{th}$ January, 2020. Number of cases increased to 3 by $3^{rd}$ February, all of them were recovered by $20^{th}$ February. There was no new case of infection till $2^{nd}$ March, after which number of cases started increasing rapidly. Therefore, $0^{th}$ day in SFEIR model is taken as $3^{rd}$ March, 2020.

Time series for active cases was taken from \cite{covid19}. Model was fitted to data using non-linear square fit method. Differential equations were solved using RK4 method \cite{singh2015modelling} for time period of 500 days with step size of 0.05. Age structure population data was taken from \cite{population}. Values of $\rho_{im}$ and $\sigma_{im}$ were taken from references mentioned in Table \ref{Table1}. All simulations were performed in python programming language.

\section{Results}

\subsection{Comparing the three variants of SFEIR model}
As discussed in the Methods section, whole population of India was divided into 16 classes of different age groups. Simulations were performed for three cases, namely;
\newline
(I) without lockdown and without fear,  \newline
(II) with lockdown (for fixed number of days) but without fear, and, \newline
(III) with lockdown (for fixed number of days) and with fear.  \newline

\begin{figure}[H]
\begin{subfigure}{0.52\textwidth}
  \centering
  \includegraphics[width=1.02\linewidth]{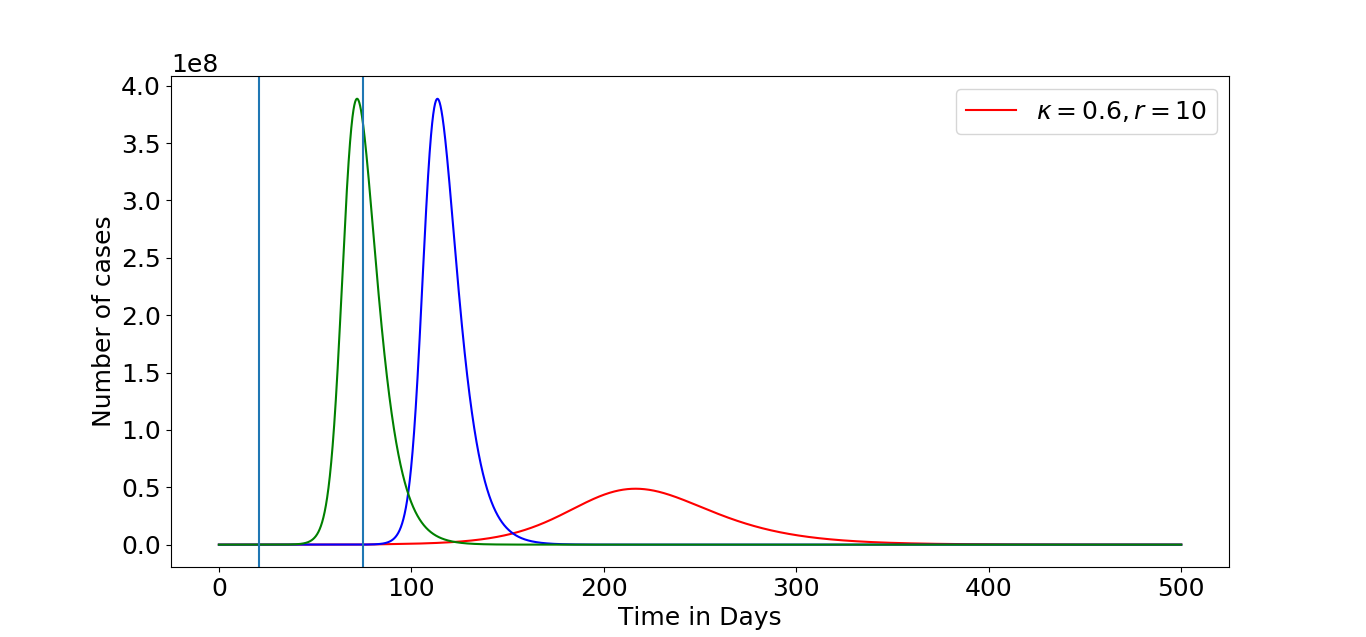}
  \caption{}
  \label{F2a}
  \end{subfigure}
\begin{subfigure}{0.52\textwidth}
  \centering
  \includegraphics[width=1.02\linewidth]{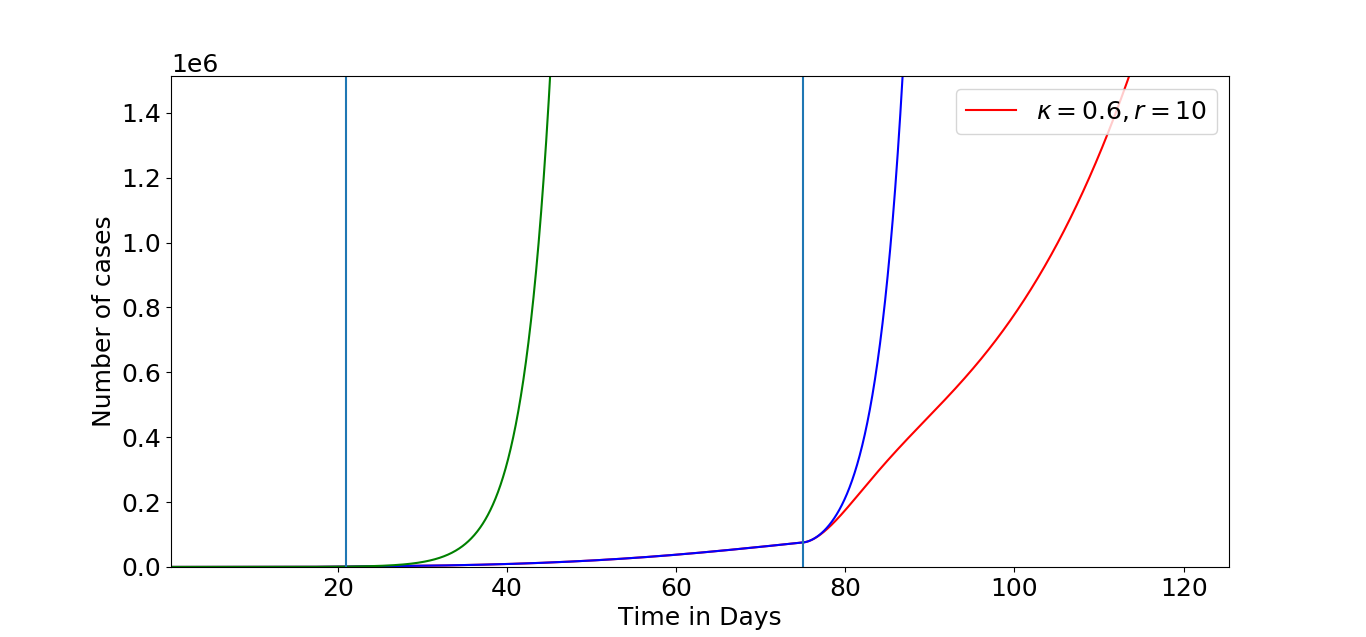}
  \caption{}
  \label{F2b}
 \end{subfigure} 
\caption{\textbf{Effect of fear conditioning on infection dynamics in post-lockdown phase:} (a) Comparative simulation results among three cases, with x-axis representing the number of days starting $3^{rd}$ March and y-axis representing the number of infected individuals. Green colored curve represents the scenario of simple SEIR model in the absence of lockdown, blue colored curve shows SEIR model with lockdown implemented from day $21^{st}$ to day $75^{th}$ and red colored curve represents the SFEIR model results with fear conditioning being implemented at $75^{th}$ day. (b) Zoomed-in portrayal of the results of three cases in smaller time scale with same color coding. Values of $\kappa$ and $r$ are kept 0.6 and 10, respectively, in both the Figures.} 
\label{F2} 
\end{figure}

Case I is represented by green curve showing simulation results in the absence of lockdown. Case II is represented by blue curve, and shows the results of lockdown intervention starting from $21^{st}$ day (March $24^{th}$) and continuing till $75^{th}$ day ($17^{th}$ May, 2020), after which contact matrix was set to as it was before. Case III is represented by red curve, in which lockdown is implemented from $21^{st}$ day to $75^{th}$ day along with SFEIR model implementation starting from $75^{th}$ day. The simulation was done for 500 days. A comparison among these cases indicates that lockdown has reduced the number of new infections substantially, but there is an abrupt rise in number of infectious cases if lockdown is lifted at one go. By implementing the SFEIR model from $75^{th}$ day, the curve flattens considerably resulting in further decrease of number of infections and its peak shifts to about $220^{th}$ day. Results are summarized in Fig. \ref{F2}.

\subsection{Variation of parameters}

\begin{figure}[H]
\begin{subfigure}{.52\textwidth}
  \centering
  \includegraphics[width=1.02\linewidth]{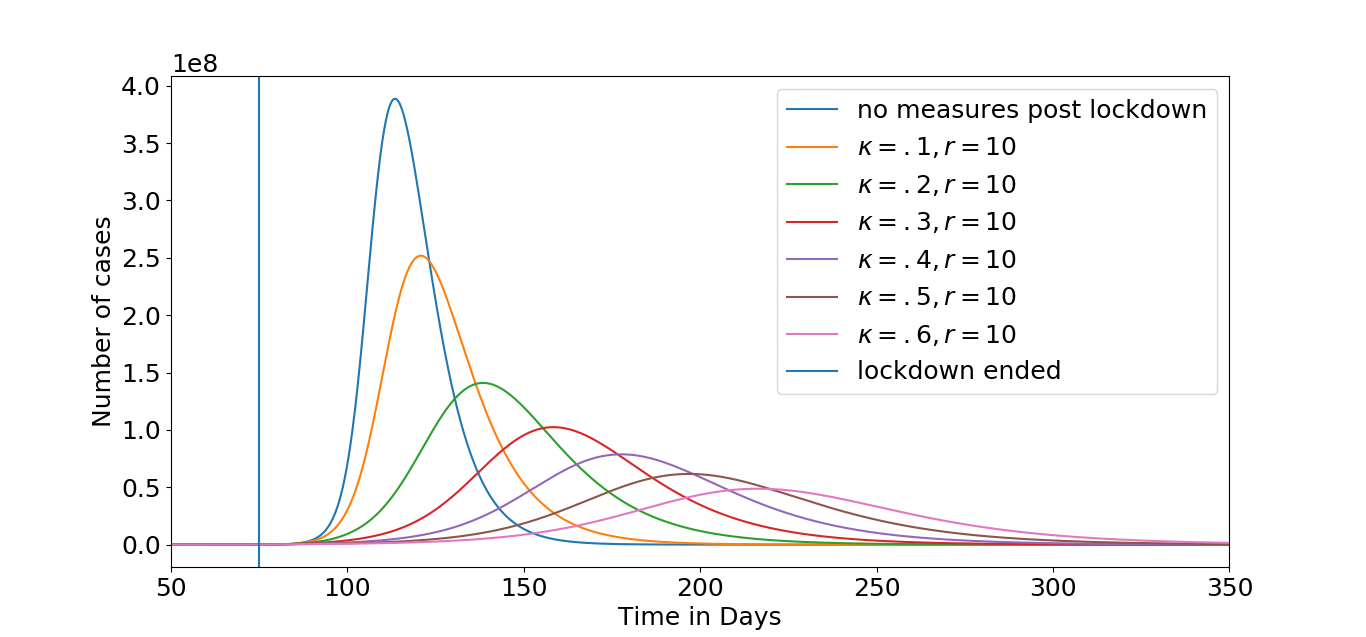}
  \caption{}
  \label{F3a}
  \end{subfigure}%
\begin{subfigure}{.52\textwidth}
  \centering
  \includegraphics[width=1.02\linewidth]{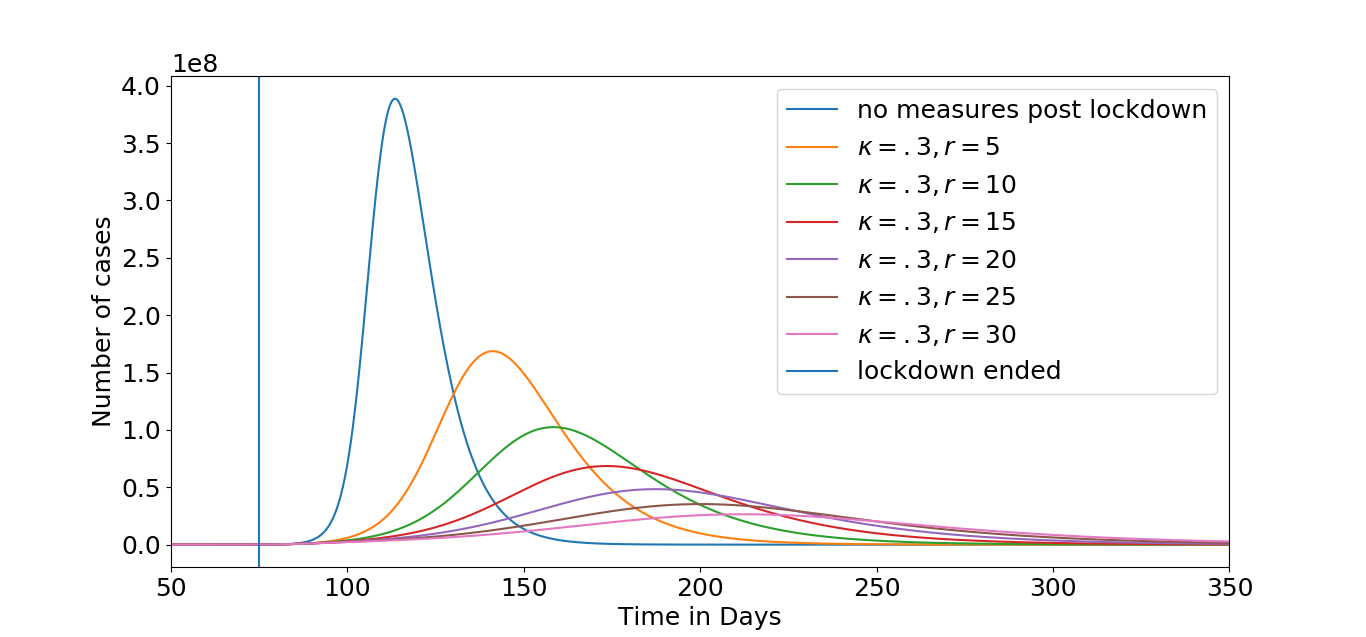}
  \caption{}
  \label{F3b}
  \end{subfigure}
\begin{subfigure}{.52\textwidth}
  \centering
  \includegraphics[width=1.02\linewidth]{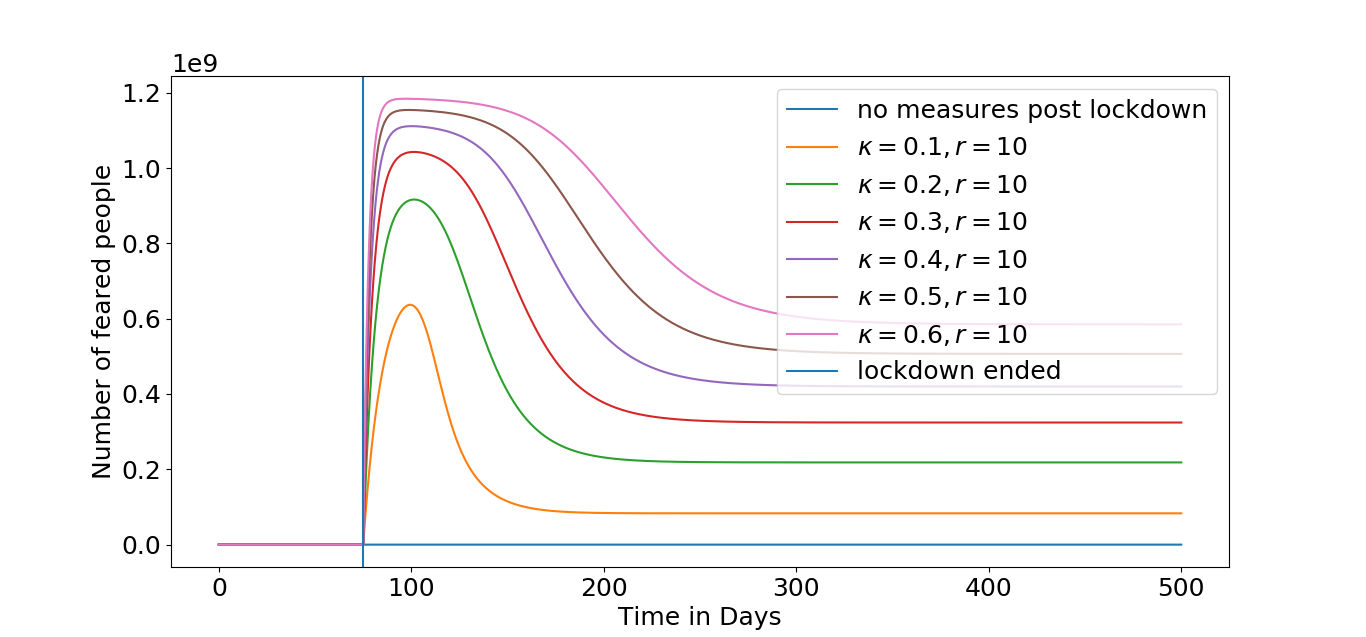}
  \caption{}
  \label{F3c}
  \end{subfigure}%
\begin{subfigure}{.52\textwidth}
  \centering
  \includegraphics[width=1.02\linewidth]{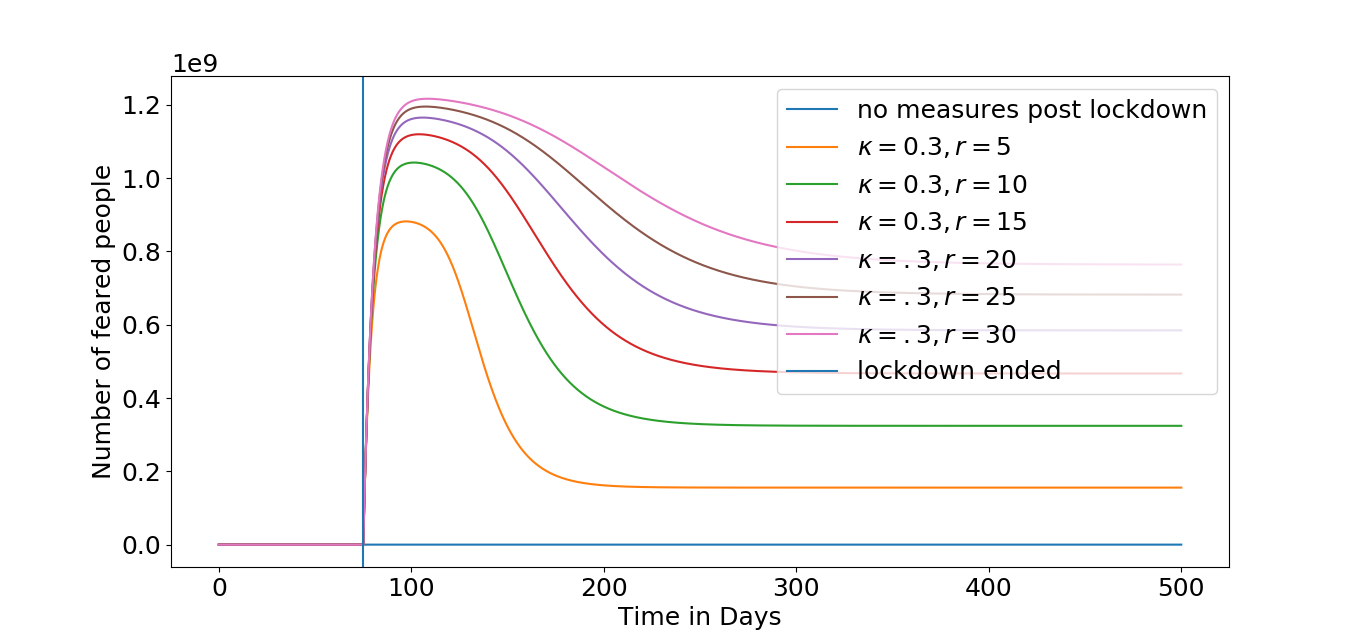}
  \caption{}
  \label{F3d}
  \end{subfigure}
\caption{\textbf{Impact of varying $\kappa$ and $r$ (regulating fear-conditioning and reward, respectively) on infection dynamics:} (a) Simulation results for different $\kappa$ values with $r$ fixed at 10 showing that the peak height is decreasing with increasing $\kappa$ values. (b) simulation results for different $r$ values with $\kappa$ fixed at 0.3. Flattening and profound shifting of curve is apparent with increase in $r$.  Increase in the number of feared people with varying the values of $\kappa$ and $r$ can be seen clearly in (c) and (d), respectively. Vertical blue line at  $75^{th}$ day represents the lifting of lockdown. $0^{th}$ day is $3^{rd}$ March, 2020.} 
\label{F3} 
\end{figure}

\subparagraph*{Effect of varying $\kappa $ and $r$ }     
Here, we have analysed the role of $\kappa$ (fraction of people who are fear conditioned by media) and $r$ (reward parameter) on the number of infected individuals. For this, we varied $\kappa$ from 0.1 to 0.6 with the step size of 0.1 and the value of $r$ was fixed at 10. Values of all other parameters were kept constant as described in Table \ref{Table1}. 
Decrease in peak height and flattening of curve is observed as shown in Fig. \ref{F3a} and \ref{F3b}. This shift in peak arises mainly due to the fact that feared population becomes infected with lesser probability than the unfeared susceptible population. Along with $\kappa$, different values of $r$ were also varied and analysed. We varied the value of $r$ from 5 to 30 with the step size of 5 and keeping the value of $\kappa$ constant at 0.3. Flattening and profound shifting of curve is again observed with the increase in $r$ values that happens mainly because increase in reward increases the time of stay in fear compartment. These results are shown in Fig. \ref{F3b}.

Effects of varying the parameters $\kappa$ and $r$ on feared population are shown in the Fig. \ref{F3c} and \ref{F3d}, respectively. An increase in values of  $\kappa$ and $r$, increases the fraction of people in feared compartment during post-lockdown phase, thereby, reducing the number of infections.

\begin{figure}[H]
\begin{subfigure}{.52\textwidth}
  \centering
  \includegraphics[width=1.0\linewidth]{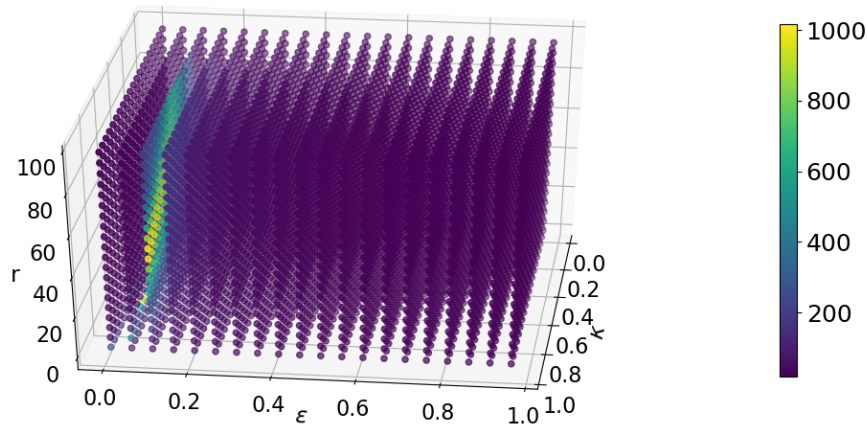}
  \caption{}
  \label{F4a}
  \end{subfigure}%
\begin{subfigure}{.52\textwidth}
  \centering
  \includegraphics[width=1.0\linewidth]{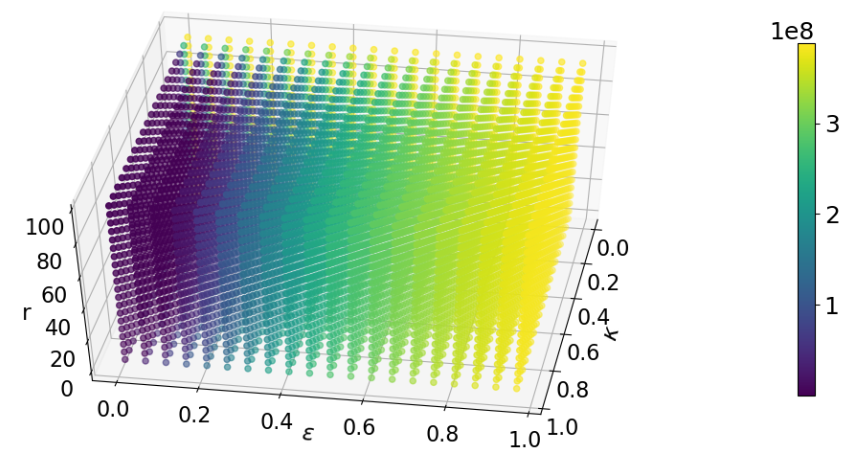}
  \caption{}
  \label{F4b}
  \end{subfigure}
\begin{subfigure}{.52\textwidth}
  \centering
  \includegraphics[width=1.0\linewidth]{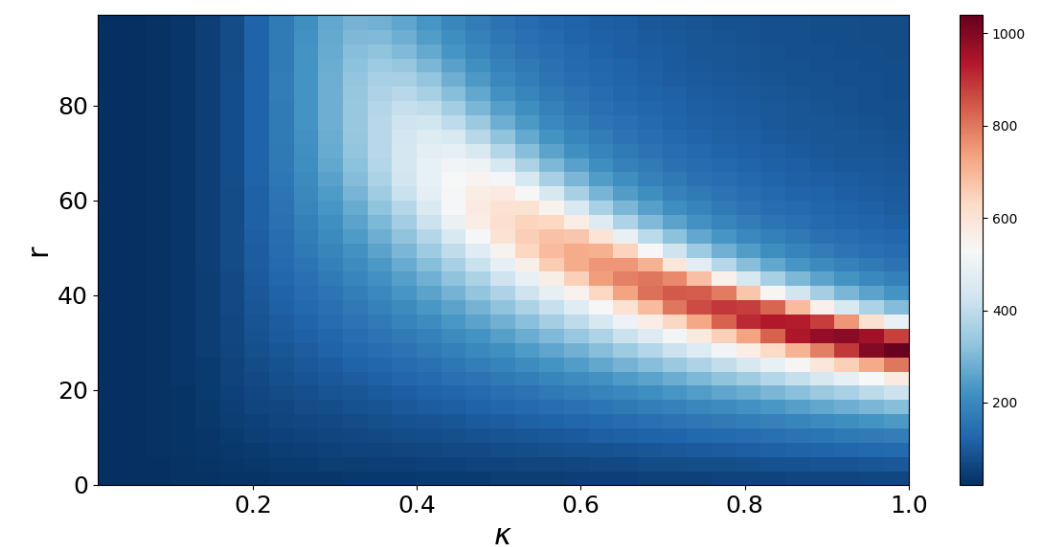}
  \caption{}
  \label{F4c}
  \end{subfigure}%
\begin{subfigure}{.52\textwidth}
  \centering
  \includegraphics[width=1.0\linewidth]{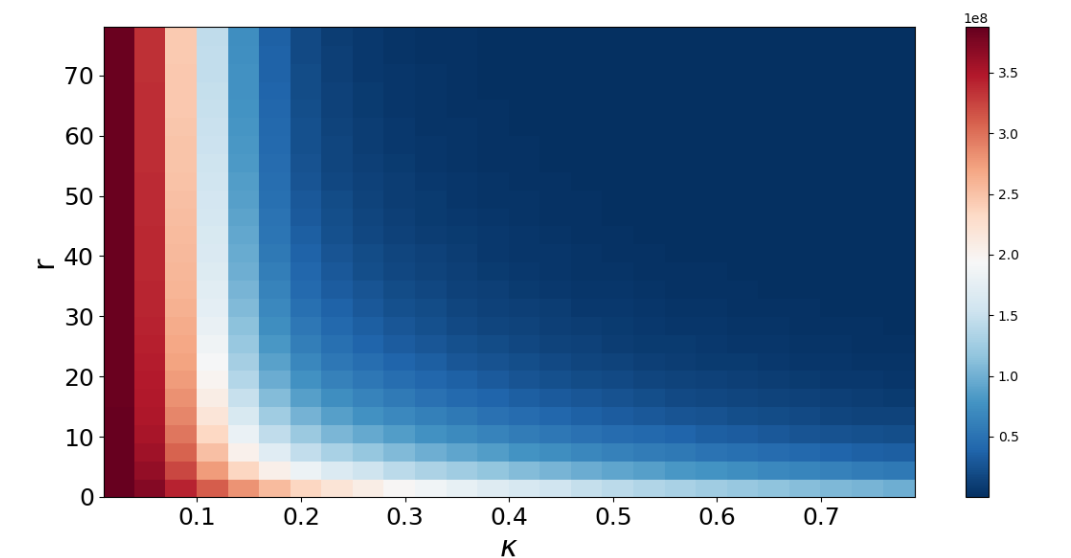}
  \caption{}
  \label{F4d}
  \end{subfigure}
\caption{\textbf{Phase space of SFEIR model:} (a) In terms of variable $W$ of infection curve, calculated as full width at half maxima of curve for that value of $\epsilon$, $\kappa$ and $r$, and (b) in terms of peak height $H$ of curve for that value of $\epsilon$, $\kappa$ and $r$. (c) and (d) represent the 2-dimensional phase space in terms of $W$ and $H$ corresponding to variables $\kappa$ and $r$. Value of $\epsilon$ was kept fixed at 0.1.} 
\label{F4} 
\end{figure}

\subparagraph{Phase space:}
To quantify the effect of simultaneously varying $\kappa$, $r$ and $\epsilon$ on infection dynamics, two phase diagrams corresponding to variables $W$  and $H$  are plotted as shown in Fig. \ref{F4a} and Fig. \ref{F4b}, respectively. Variables $W$  and $H$ are defined as,

\begin{equation}
\begin{aligned}
 W &= {Full\; width\; at\; half\; maxima} \\
 H &= {Peak\; height\; of\; curve}
\end{aligned}
\end{equation}

$\kappa$ and $\epsilon$ values were varied from 0.01 to 1 in the steps of 0.05, and $r$ values were varied from 1 to 100 in the steps of 5. By looking at plots, it can be seen that $W$ divides the phase space into 3 parts: Region I, II and III. Region II corresponds to the large values of variable $W$ represented as yellow-green spheres in Fig. \ref{F4a}. For the corresponding values of $\kappa$, $r$ and $\epsilon$ in Fig. \ref{F4b}, variable $H$ has moderate values. 

On both sides of region II in Fig. \ref{F4a}, variable $W$ has small values. In the left side part of region II in Fig. \ref{F4a}, variable $W$ has small values and for the corresponding values of $\kappa$, $r$ and $\epsilon$ in Fig. \ref{F4b}, variable $H$ is also small. This part of phase space is termed as Region I. While in the right side part of region II in Fig. \ref{F4a}, variable $W$ has small values and for the corresponding values of $\kappa$, $r$ and $\epsilon$ in Fig. \ref{F4b}, variable $H$ is high. This part of phase space is termed as Region III.

Thus, the three regions (Region I, II and III) are defined on the basis of values of $W$ and $H$ as follows: 
\newline
Region I: Small $W$, small $H$   \newline
Region II: Large $W$, moderate $H$   \newline
Region III: Small $W$, large $H$

Region I (small $W$ values, small $H$ values) corresponds to conditions in which number of infections are very less and pandemic ends swiftly. Region II corresponds to conditions under which rise in number of cases is slow (curve is flat), but it takes more time for pandemic to mitigate. Region III corresponds to high severity of pandemic. Although, Region I looks desirable but it can not be reached as it corresponds to very less number of infections (in the range of ${10}^5$ only) in India. One has to focus on developing strategies to avoid reaching into the Region III, and to remain particularly in Region II. 

As region II corresponds to value of $\epsilon$ in range 0.05 to 0.15. For the rest of study, $\epsilon$ was fixed to 0.1. To ascertain the effect of $r$ and $\kappa$ on infection dynamics, 2-D version of Fig. \ref{F4a} and \ref{F4b} are plotted in Fig. \ref{F4c} and \ref{F4d}, respectively. The three regions can be clearly seen in Fig. \ref{F4c}. The red-white curve has large $W$ values and corresponds to region II, the upper-right blue quadrant corresponds to region I and lower-left blue quadrant corresponds to region III of Fig. \ref{F4a}.

\subsection{Effect of keeping schools closed in post-lockdown phase}
\begin{figure}[H]
\begin{subfigure}{0.52\textwidth}
  \centering
  \includegraphics[width=1.02\linewidth]{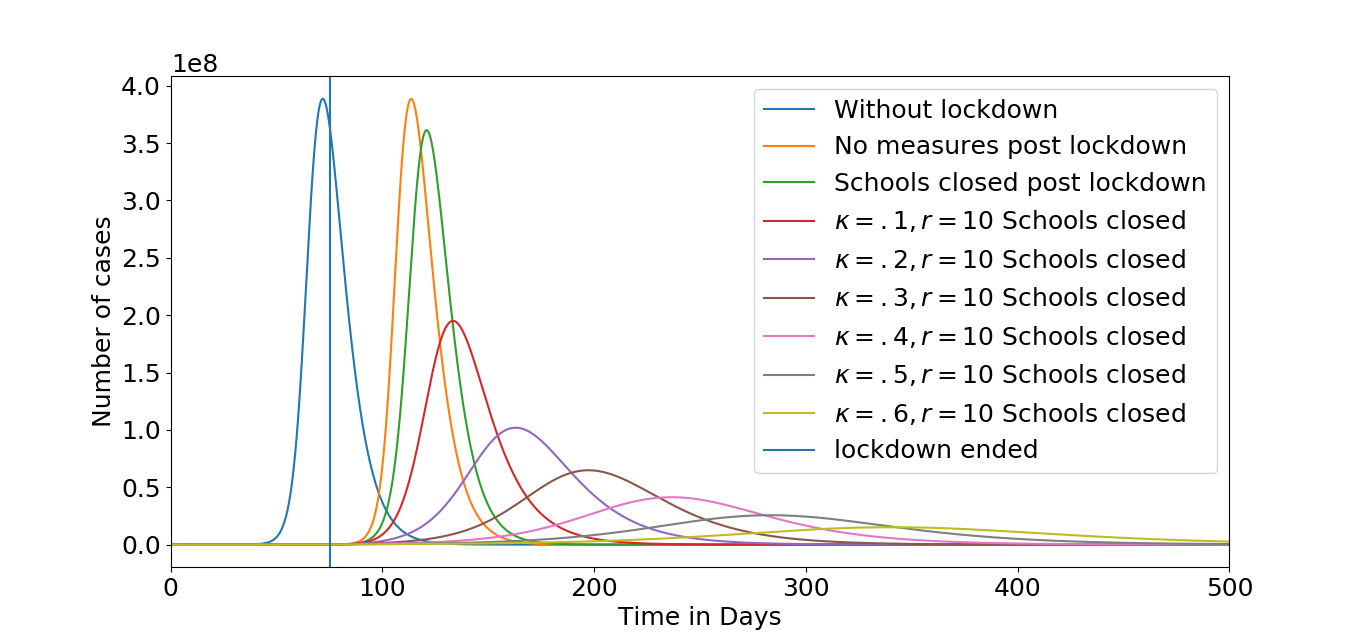}
  \caption{}
  \label{F5a}
  \end{subfigure}%
\begin{subfigure}{0.52\textwidth}
  \centering
  \includegraphics[width=1.02\linewidth]{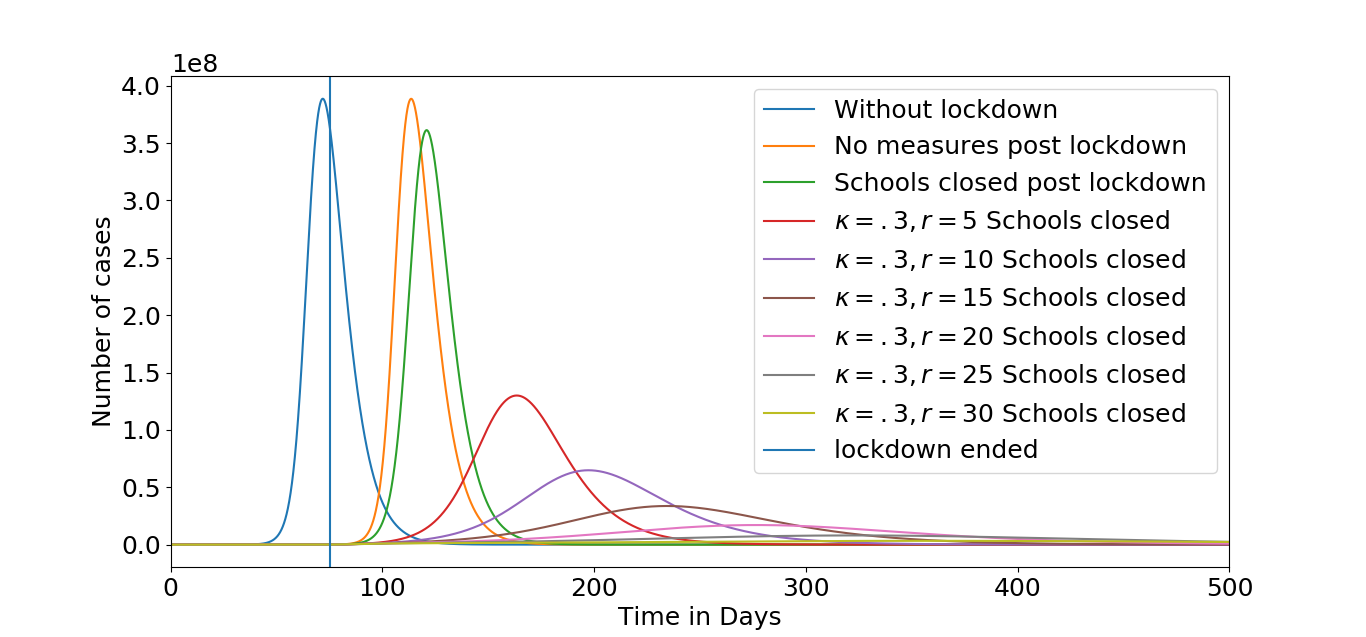}
   \caption{}
  \label{F5b}
  \end{subfigure}
\begin{subfigure}{0.52\textwidth}
  \centering
  \includegraphics[width=1.02\linewidth]{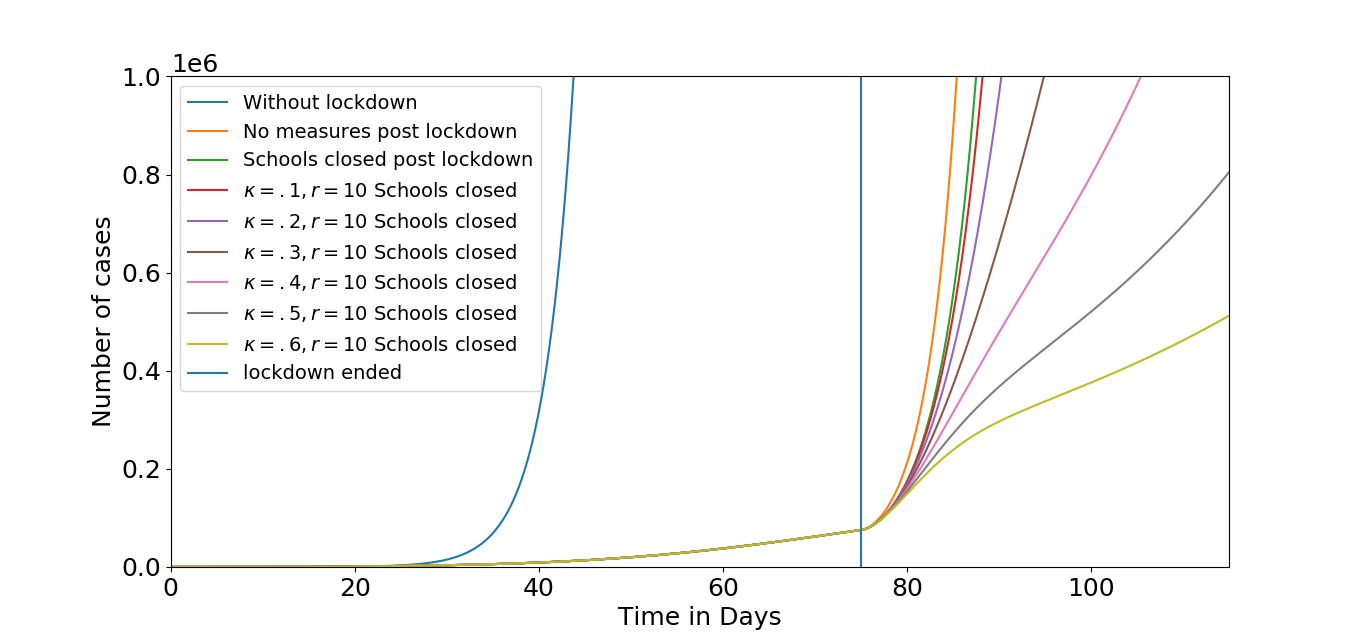}
  \caption{}
  \label{F5c}
  \end{subfigure}%
\begin{subfigure}{0.52\textwidth}
  \centering
  \includegraphics[width=1.02\linewidth]{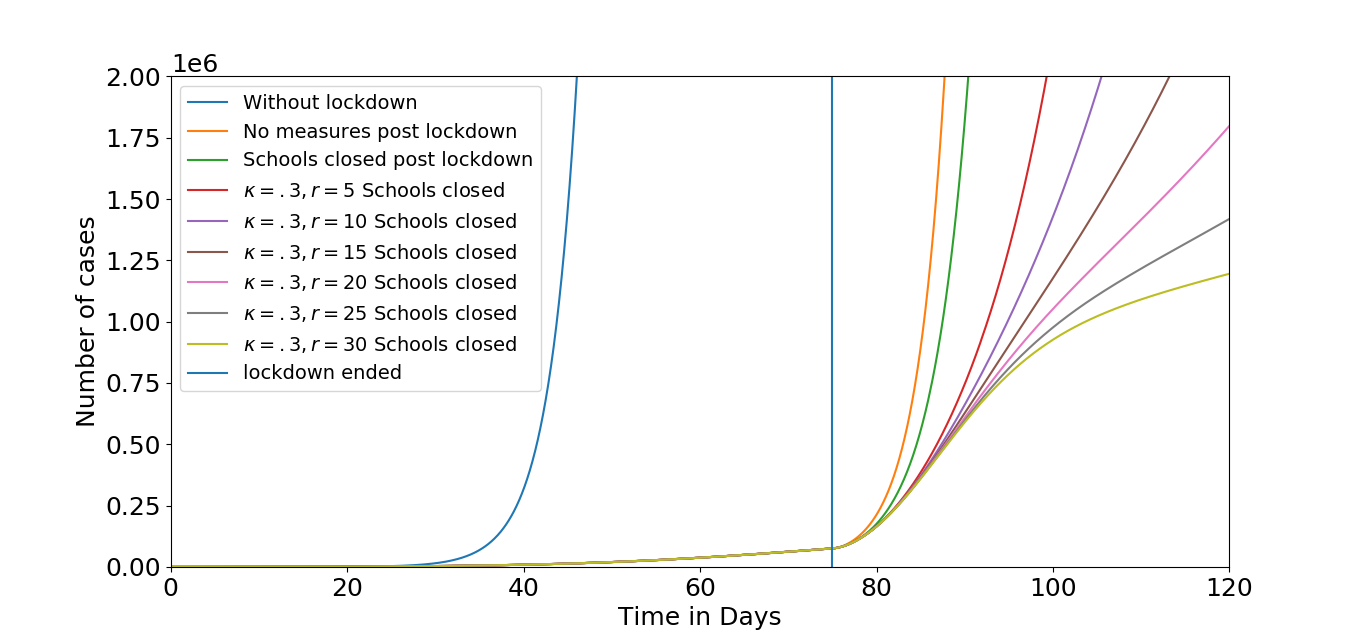}
   \caption{}
  \label{F5d}
  \end{subfigure}
\caption{\textbf{Forecast of post-lockdown epidemic progression in schema of fear conditioning with closed schools:} While keeping the schools closed, (a) simulation results for different $\kappa$ values with $r$ fixed at 10 and (b) simulation results for different $r$ values with $\kappa$ fixed at 0.3. Both (c) and (d) shows the smaller time plot of (a) and (b), respectively. $0^{th}$ day is $3^{rd}$ March and vertical blue line at $75^{th}$ day ($17^{th}$ May) signifies the day of lockdown ending.} 
\label{F5} 
\end{figure}

Closure of schools is one of the most often considered non-pharmaceutical intervention to control a pandemic \cite{cauchemez2008estimating}. In this section, we first analyse the effect of keeping schools closed after the lockdown is over \textit{i.e.} on $75^{th}$ day ($17^{th}$ May in present scenario), in the absence of fear. This is done by keeping the values of contact matrix of school, $C_{ij}^{S} = 0$. As seen in Fig. \ref{F5a}, this step, while shifting the peak of the curve, reduces the height of peak very little ascompared to original curve. Hence, it may be concluded that by just keeping schools closed and opening all other places after the lockdown will not have substantial effect on the containment of epidemic.

In order to decipher the effect of media induced fear, we simulated the effect of closing schools in the proposed SFEIR model and observed a drastic reduction in number of infected individuals. The curve becomes considerably flat.

\subparagraph{Variation of $\kappa$}
We tried to analyse the effect of varying $\kappa$ while keeping the schools closed. We first kept $r$ fixed at value of 10 and varied $\kappa$ from 0.1 to 0.6 with step size of 0.1. As the value of $\kappa$ increases, curve almost becomes flat with curve peak height at around 175 days for $\kappa$ =0.2 (shown in Fig. \ref{F5a}). The peak height of curve shifts further to around 200 days with further flattening as $\kappa$ increases to 0.3. Thus, keeping schools closed in post-lockdown simultaneously with the fear induced conditional learning can have drastic effect in reducing the number of infected individuals. Fig. \ref{F5c} shows the smaller time frame of Fig. \ref{F5a}, in which it can be clearly seen that the slope of curve decreases sharply as the value of $\kappa$ increases. 

\subparagraph{Variation of $r$}
To see the effect of $r$ in conjunction with keeping schools closed, we kept $\kappa$ at a constant value of 0.3 and varied $r$ from 5 to 30 with the step size of 5. The curve peak for $r$ = 5 shifts to $180^{th}$ day and raising $r$ to 10 further shifts the peak to around $200^{th}$  day as well as flattens it significantly as shown in Fig. \ref{F5b}. In smaller time frame, in Fig. \ref{F5d}, it can be easily seen that as $r$ increases, slope of the curve decreases. 

\subsection{Comparative analysis of fear conditioning with schools closure for different time periods}

\begin{figure}[H]
\begin{subfigure}{0.52\textwidth}
  \centering
  \includegraphics[width=1.02\linewidth]{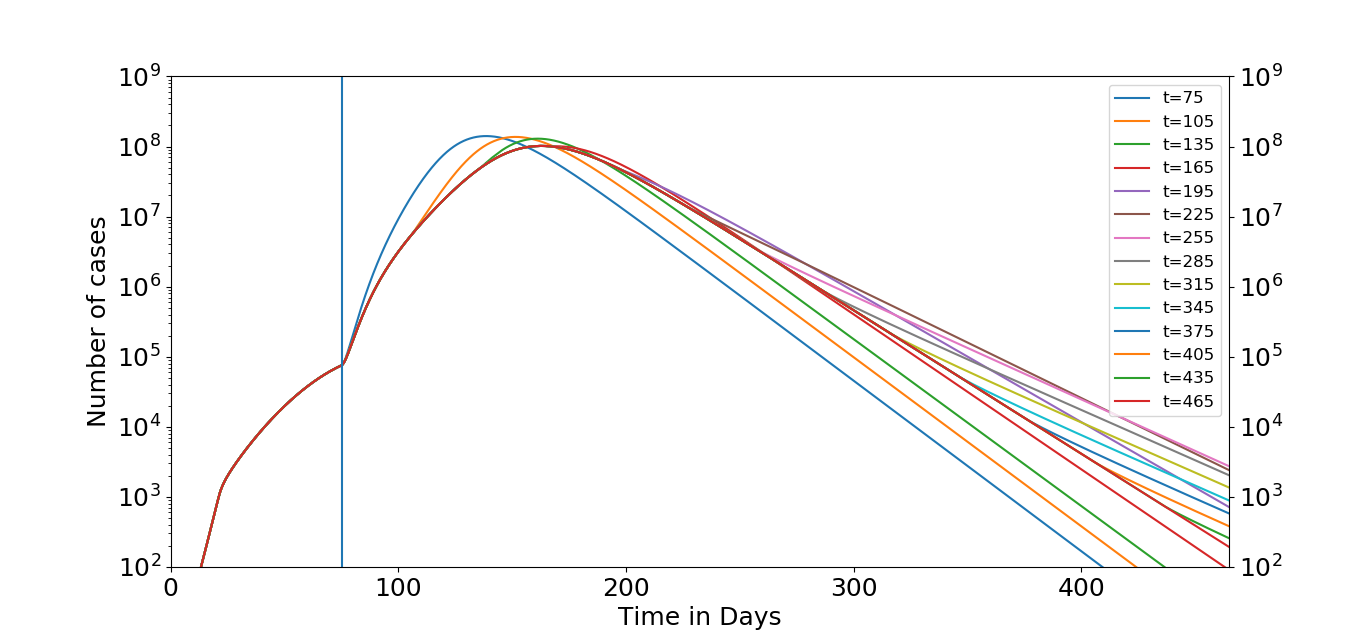}
  \caption{}
  \label{F6a}
  \end{subfigure}%
\begin{subfigure}{0.52\textwidth}
  \centering
  \includegraphics[width=1.02\linewidth]{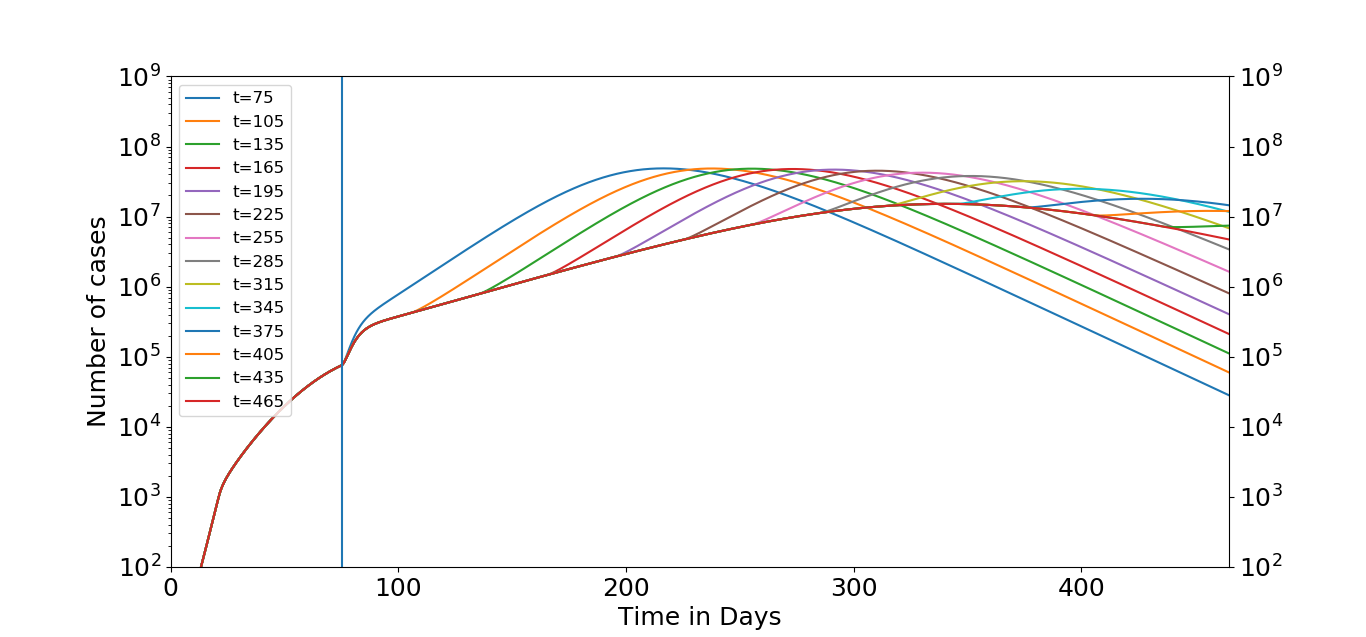}
  \caption{}
  \label{F6b}
  \end{subfigure}
\begin{subfigure}{0.52\textwidth}
  \centering
  \includegraphics[width=1.02\linewidth]{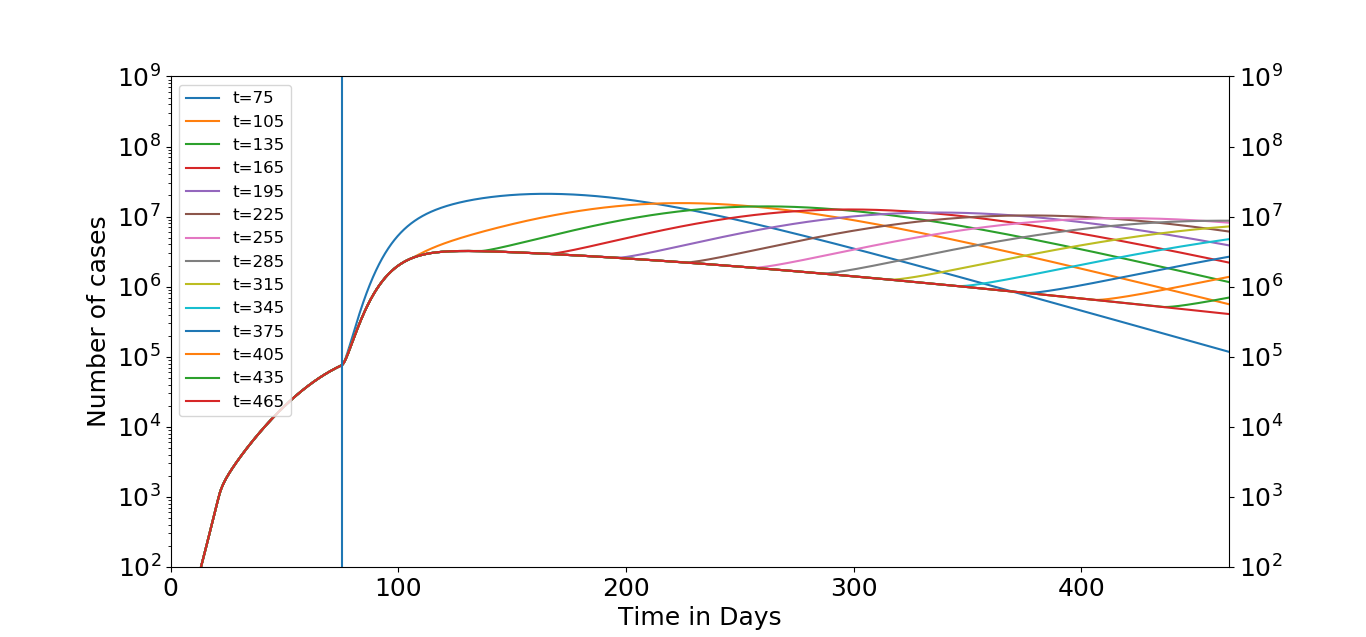}
  \caption{}
  \label{F6c}
  \end{subfigure}%
\begin{subfigure}{0.52\textwidth}
  \centering
  \includegraphics[width=1.02\linewidth]{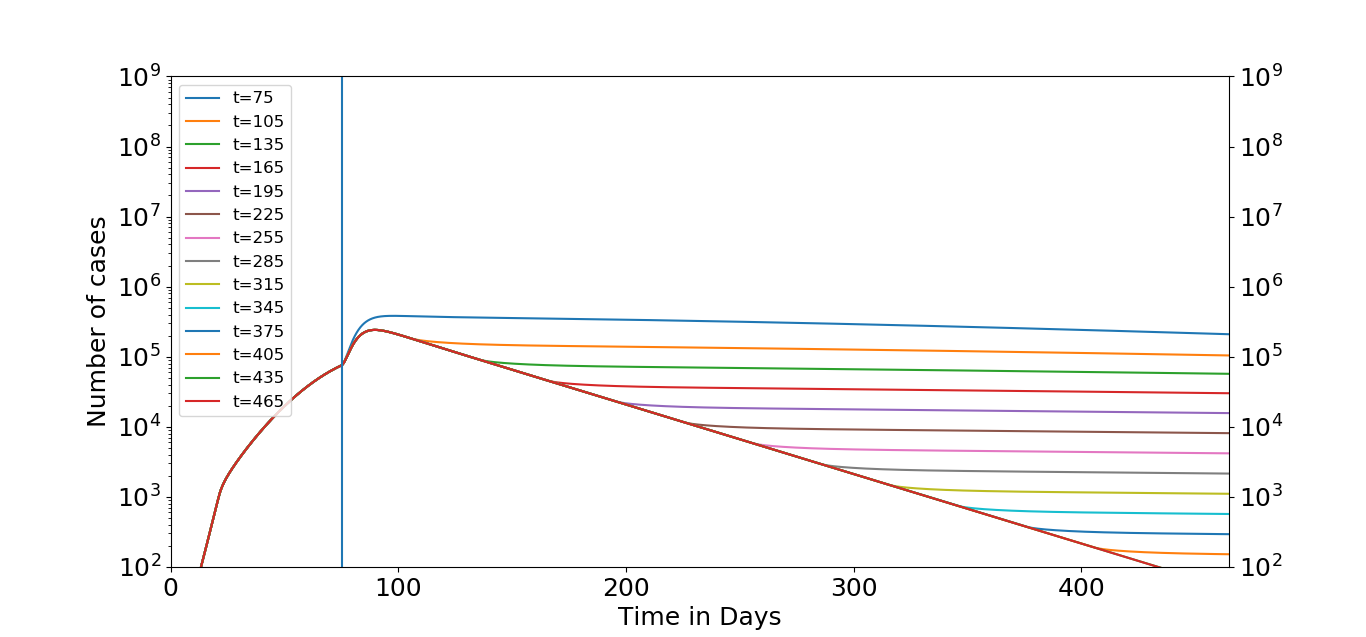} 
  \caption{}
  \label{F6d}
  \end{subfigure}
\caption{ \textbf{Comparative simulations of schools closure duration and fear conditioning intensity for different time periods}. Simulation results show when $\kappa$ and $r$ were kept at 0.2 and 10 in (a), $\kappa$ and $r$ were kept at 0.6 and 10 in (b), $\kappa$ and $r$ were kept at 0.2 and 60 in (c) and $\kappa$ and $r$ were kept at 0.6 and 60 in (d). $0^{th}$ day is $3^{rd}$ March and $75^{th}$ day ($17^{th}$ May) signifies the day of lockdown ending. In each panel, differently colored trajectories correspond to the infection dynamics on that day of schools opening. It can be seen in panel (d) that closing schools upto 420 days can bring the number of infections below 100.}
\label{F6}
\end{figure}

Here, we have analysed the time needed to keep schools closed along with the intensity of fear conditioning (characterized by $\kappa$ and $r$) required for controlling the COVID-19 pandemic in post-lockdown phase in India. We chose two set of values of $\kappa$ and $r$ from Fig. \ref{F3c} and \ref{F3d} one belonging to high peak region and and the other belonging to low peak region. $\epsilon$ was kept fixed to 0.1. Thus, four cases were designed:
\vspace{2mm}\newline
Case I (Fig. \ref{F6a}) :  $\kappa$ (high peak region) = 0.2 and $r$ (high peak region) = 10   \newline
Case II (Fig. \ref{F6b}) : $\kappa$ (low peak region) = 0.6 and $r$ (high peak region) = 10   \newline
Case III (Fig. \ref{F6c}) : $\kappa$ (high peak region) = 0.2 and $r$ (low peak region) = 60   \newline
Case IV (Fig. \ref{F6d}) : $\kappa$  (low peak region) = 0.6 and $r$ (low peak region) = 60   \newline

For each case, different time periods of keeping schools closed were plotted starting from 75 days (\textit{i.e.}  $17^{th}$ May, in present scenario) to 465 days with step size of 30 days.
In case I, severity of pandemic is still substantial and schools closure does not have much effect.
In cases II and III, if schools are kept closed, exponential increase in number of infections is halted. This will constraint the growth of pandemic but as soon as schools are repopened, number of infections will spike upward. Therefore, schools need to be kept closed for very long period of time, which is not at all practical. However, this combination can control the epidemic spread till the vaccines and medicines are available. Results are summarized in Fig. \ref{F6b}  and \ref{F6c} . In case IV, as shown in Fig. \ref{F6d}, the curve has a negative slope if schools are kept closed. The number of cases reduces to less than 100 if schools are opened after about 420 days from $3^{rd}$ March, 2020.
Therefore, $\kappa$ and $r$ values both in low peak region can have a significant effect in constraining the pandemic.

\subsection{Effect on infections due to fear conditioning on specific age-groups}

\begin{figure}[H]
\begin{subfigure}{0.52\textwidth}
  \centering
  \includegraphics[width=1.02\linewidth]{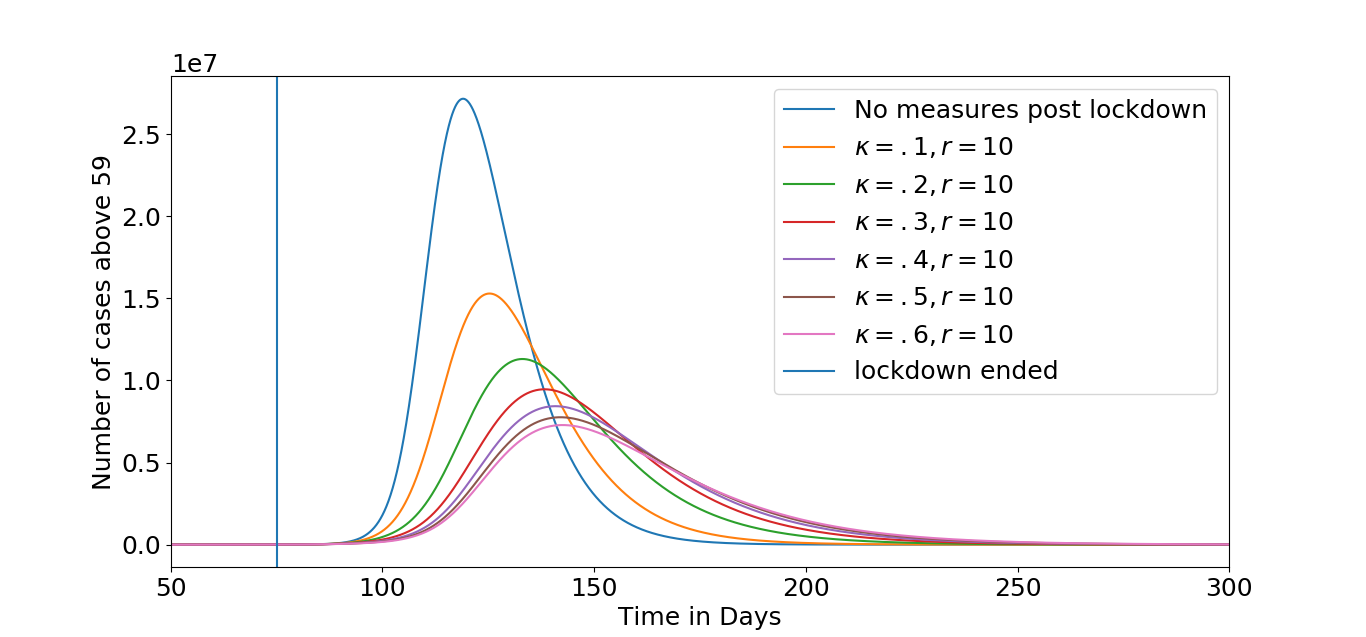}
  \caption{}
  \label{F7a}
  \end{subfigure}%
\begin{subfigure}{0.52\textwidth}
  \centering
  \includegraphics[width=1.02\linewidth]{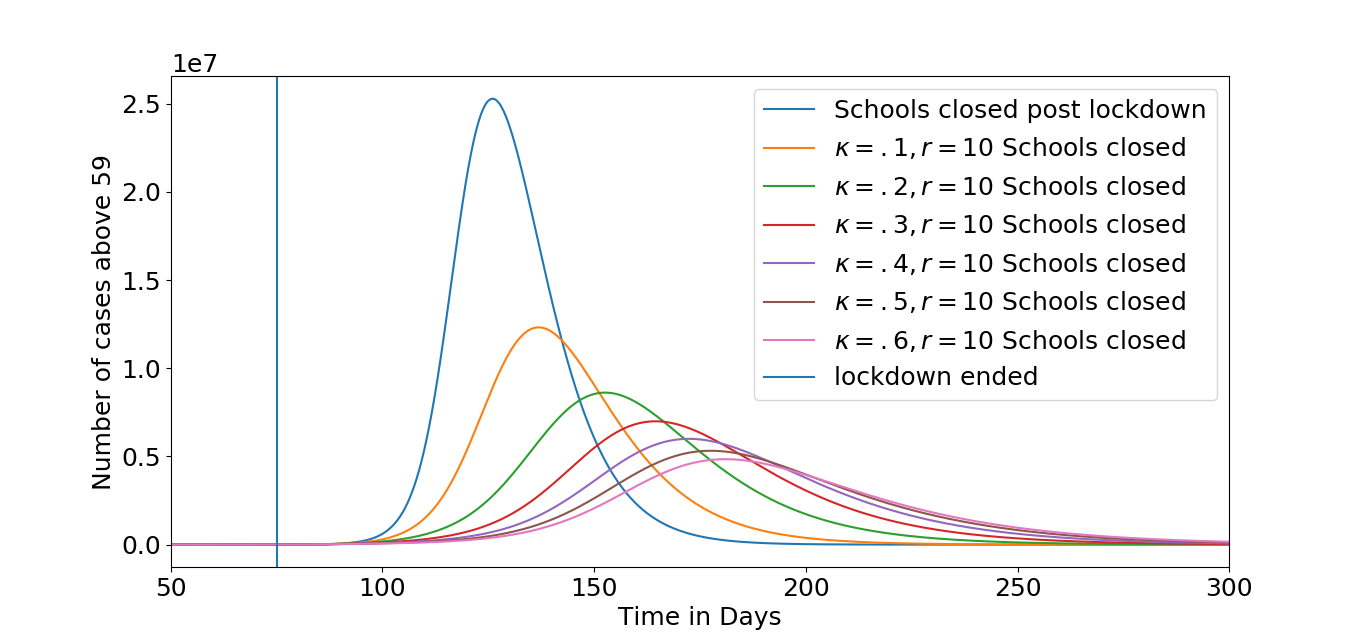}
  \caption{}
  \label{F7b}
  \end{subfigure}
\begin{subfigure}{0.52\textwidth}
  \centering
  \includegraphics[width=1.02\linewidth]{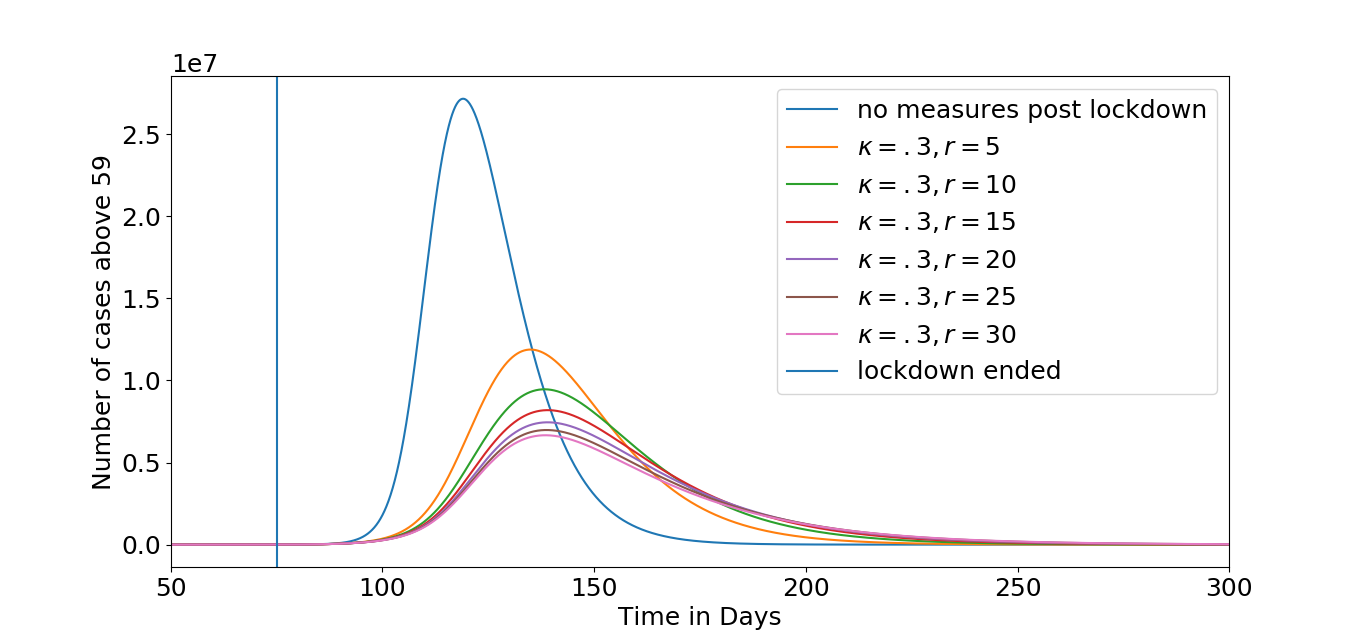}
  \caption{}
  \label{F7c}
  \end{subfigure}%
\begin{subfigure}{0.52\textwidth}
  \centering
  \includegraphics[width=1.02\linewidth]{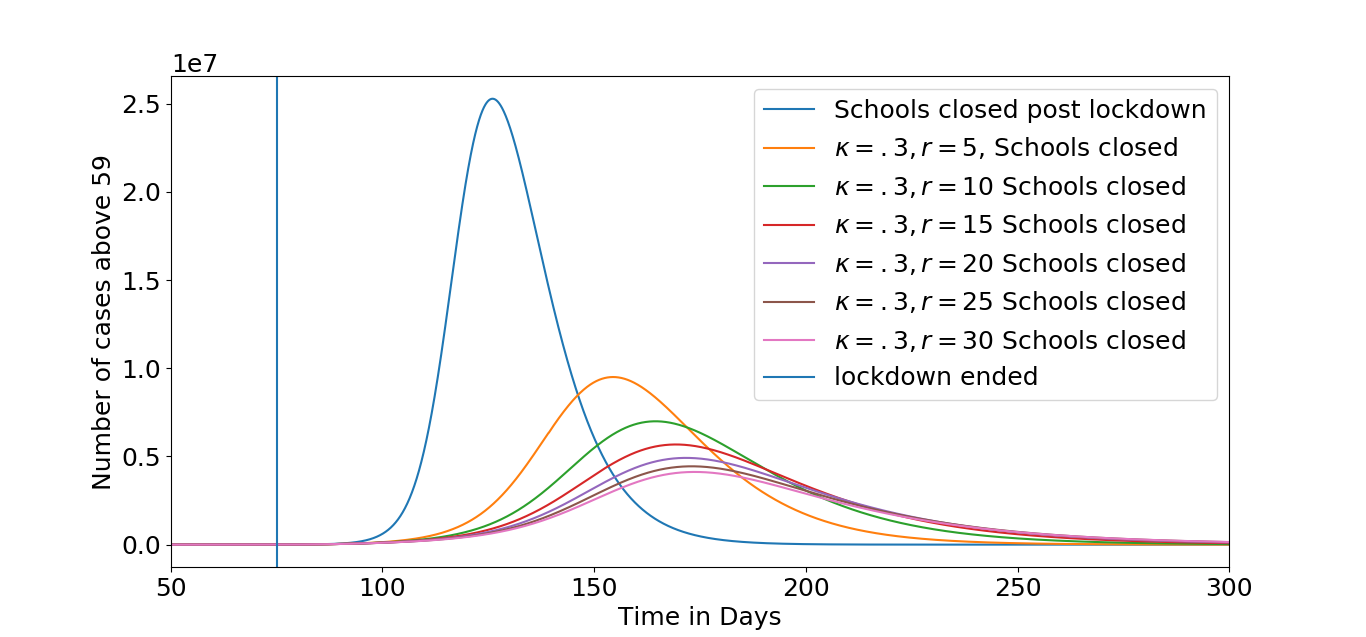} 
  \caption{}
  \label{F7d}
  \end{subfigure}
\caption{\textbf{Infections in old age people when middle aged are fear conditioned:} (a) Simulation results of varying $\kappa$ in $G_2$ on infections in $G_3$, $r$ was kept fixed at 10 in $G_2$ and $G_3$ and 0 in $G_1$, $\kappa$ was kept fixed at 0 in $G_1$ and 0.1 in $G_3$. 
 (b) Effect of keeping schools closed and varying $\kappa$ in $G_2$ on population in $G_3$. All other parameters were kept same as in (a). (c) Simulation results of varying $r$ in $G_2$ on infections in $G_3$ , $\kappa$ was kept fixed at 0.1 in $G_2$ and $G_3$ and 0 in $G_1$, $r$ was kept fixed at 0 in $G_1$ and 10 in $G_3$. (d) Effect of keeping schools closed and varying $r$ in $G_2$ on population in $G_3$. All other parameters were kept same as in (b).}
\label{F7}
\end{figure}

\subparagraph*{Effect of varying $\kappa$ among young and middle age population on old age population}

Across the world, critical care admission and mortality rate have been highest in people with age greater than 60 \cite{onder2020case, cdc, who, dowd2020demographic}. Hence, we checked what will be the effect of increasing fear conditioning among young and middle aged people on the infections in old individuals.
For this, we divided the number of infected individuals into three groups: $G_1$, 0- 9 years old;  $G_2$, 10-59 years old; $G_3$, 60-79 years old. 

To realise the same, we fixed the values of $\kappa$ to $0$ for $G_1$ and $0.1$ for the $G_3$, and varied $\kappa$ for $G_2$ keeping the reward $r$ equal to $0$ for group $G_1$ (as $\kappa$ is zero for this group, it doesn’t matter what value of $r$ is used) and $10$ for other two groups. 

In  Fig. \ref{F7a}, it can be seen that increasing the $\kappa$ in $G_2$ reduces the number of infections in $G_3$. If schools are also kept closed, infection numbers reduced further as shown in \ref{F7b}. Thus, keeping schools closed and inducing information based fear conditioning among young and middle-age group population can significantly reduce the infections among older population in post-lockdown period. Reduction in peak height and flattening of curve expound the effect of fear.

\subparagraph*{Effect of varying $r$ among young and middle age population  on old age population} 
Lastly, effect of varying reward $r$ among young and middle age group was evaluated on infections among old age people. $\kappa$ for $G_1$ was kept at 0, for $G_3$ at 0.1 and for $G_2$ at 0.3. Reward parameter $r$ was kept 0 for $G_1$, 10 for $G_3$ and was varied from 5 to 30 in steps of 5 for $G_2$. Number of infectious individuals among members of $G_3$ reduces as seen in  Fig. \ref{F7c} and \ref{F7d}. When supplemented with keeping schools closed in post lock-down phase, number of infections among $G_3$ were observed to be further reduced, as shown in  Fig. \ref{F7d}.

\section{Discussion}
Fear is an emotion that plays an important role in the survival of an individual. Mass media (like, television, community radio, internet and print media etc.), having ability to disseminate the information to a lot of people even in remote areas, can also very easily propagate the fear about a disease among masses. Thus, its proper usage can affect the course of progression of COVID-19 as was observed during polio vaccination programs \cite{obregon2009achieving}. To highlight the importance of media in order to make community driven changes rather than forced interventions from government, in this work, we modelled the media induced fear conditioning supplemented by positive reinforcement in the Indian population. 
Reward parameter $r$, quantifying the effect of positive reinforcement \textit{via} various media platforms, was used as an abstract parameter that can represent various methods of positive reinforcement, like, praise for having certain beliefs that may be accorded by the social, religious and political leaders. The simulation results show that if more people become fearful, the number of COVID-19 infections will decrease significantly. This manifests mainly because fearful people will socialize less \cite{felix2016bidirectional}, and will take more precautions in their daily life. Excessive fear conditioning can result in abnormal behaviours, like, specific phobias \cite{stein2006specific} and obsessive compulsive disorder \cite{milad2013deficits}. Therefore, care must be taken while developing practical methods of implementing fear conditioning so as to minimize the risk of such abnormal behaviors.

The SFEIR model also predicted that only keeping schools closed in post-lockdown phase while opening everything else will not substantially reduce the total number of infections, a finding that has also been reported in \cite{viner2020school}. However, SFIER model suggests that if paired with moderately high fear conditioning, schools closure can have a significant effect on mitigating COVID-19 spread. 
This can be explained by the fact that keeping the schools closed decreases the number of contacts significantly for younger generation aged 0-30 \cite{prem2017projecting} but does not alter the probability of infection. Fear conditioning on the other hand works by reducing the probability of transmission as well as the number of contacts. Therefore, when two of these are used simultaneously, probability of infection transmission decreases significantly.

\begin{figure}[!htb]
\centering
\includegraphics[height=7cm, width=10.5cm]{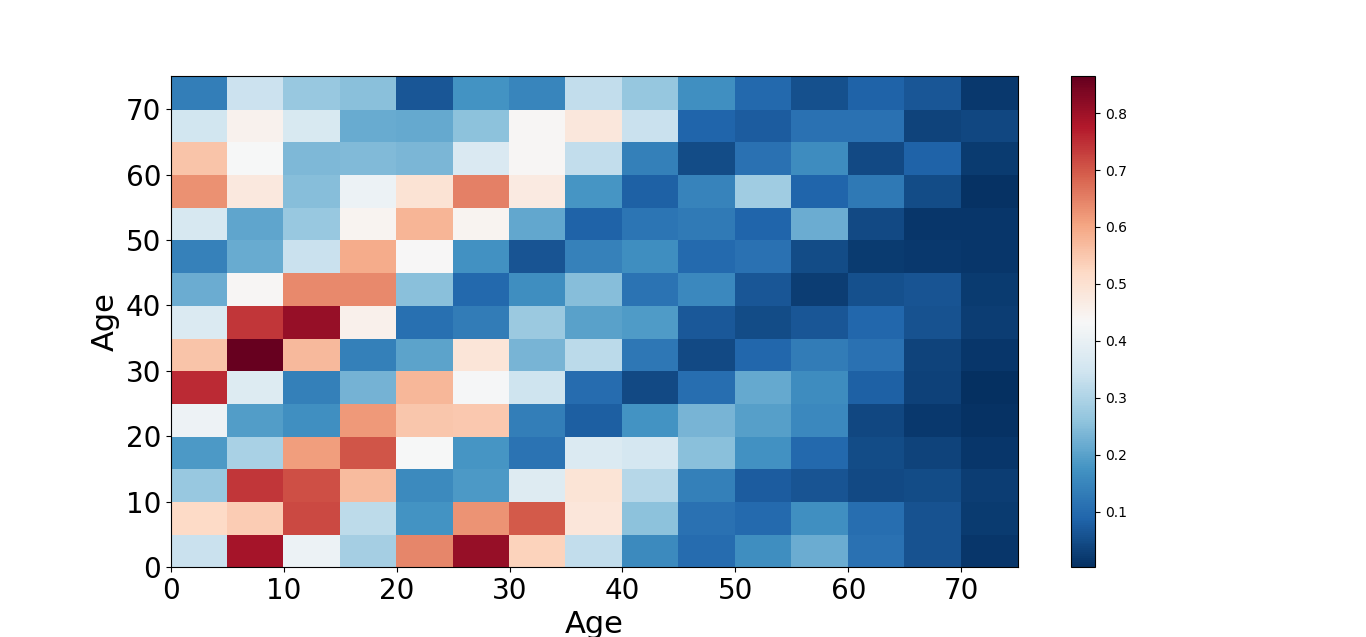}
\caption{\textbf{Age structured home interactions in India.} At home, people of all generations interact with each other in India. Data taken from \cite{prem2017projecting}.}
\label{F8}
\end{figure}

Against COVID-19 pandemic, people in the age group $G_3$ have been found to have the highest propensity of being admitted to ICUs and also have the highest mortality rate \cite{onder2020case}. One of the important predictions of the model is that if people in age-group 10-59 ($G_2$) are fearful, this will have a substantial effect on number of infections in age group 60-79 ($G_3$). In India, people of all the generations interact with each other within home (as depicted in Fig. \ref{F8}). So, if people in age group (10-59) are sufficiently fear conditioned, the probability of maintaining physical distance among these two groups will increase leading to effectively decreasing the number of infections in older generation even if older generation is not highly fear conditioned. In conjunction to this, keeping the schools closed in the post-lockdown phase can further reduce the number of infections among group $G_3$. This can be explained by the fact that India has higher probability of contact between people in age group 0-29 and age group 60-79 as seen for the contact data at home shown in  Fig. \ref{F8} below. 

\subparagraph{Limitations of the present study:}
First, lack of proper age-structured data on media consumption limits the scope of model simulation. Second, standard deviation in fear time can be huge as it is also dependent upon personality of an individual. Third, reward parameter, $r$ and fraction of people developing fear $\kappa$ both should ideally be time dependent, but we have taken these as constant parameters.

\section{Conclusion and policy perspectives}

To the best of our knowledge, the SFEIR model developed in this work is the first attempt to incorporate media induced fear conditioning in mitigating the COVID-19 pandemic. The model predicts that fear conditioning \textit{via} mass media might be a helpful strategy in reducing the COVID-19 infections in post-lockdown phase. Increasing fraction of people in fear compartment by classical conditioning and increasing fear time by operant conditioning reduced the number of infections significantly. If used in conjunction with keeping schools closed in post-lockdown phase, substantial reduction in number of infections was observed. We also examined the differential fear conditioning on different age groups in controlling the infections in elderly people. If fear conditioning is increased on individuals of age 10-59 years (group $G_2$) and exempted on kids of age 0-9 years (group $G_1$), number of infections in senior citizens of age 60-79 (group $G_3$), known to have highest mortality and hospitalization rate, is found to be reduced significantly. This condition of media induced fear conditioning supplemented with closed schools in post-lockdown phase can further decrease the number of infections in elderly people.  

We propose that the presented schema of media induced fear conditioning augmented with closing of schools (including colleges and universities) in post-lockdown phase can serve as an important non-pharmaceutical intervention to prevent healthcare system from getting overburdened. 

At the same time, along the lines of \cite{dhar2020critique}, we would like to state that modelling results should not be interpreted as the policy guidelines too strictly. We would like to add one more precaution that, although, it is being proposed to keep the schools and universities closed for long time alongwith media induced fear conditioning, it should not be construed as proposing  that other places of mass gatherings can be opened immediately. We would like to further state that our results may be treated as the guiding factor for making policy interventions but it does not mean to say that media should start showing COVID-19 news 24x7 hours. Of course, it will have a positive effect on smoothening the curve but maintaining a balance in media streaming is necessary so as to not induce fear psychosis in the citizens. Social, religious as well as political leaders have the capability and responsibility to supplement positive reinforcement for this media induced fear. 

\section{Declaration of Competing Interest}
The authors declared that there is no conflict of interest.

\bibliography{covid}

\begin{thebibliography}{10}

\bibitem{whocovid19}
WHO.
\newblock World health organization.
\newblock
  \url{https://www.who.int/news-room/detail/08-04-2020-who-timeline---covid-19},
  2020.
\newblock Accessed: 2020-05-14.

\bibitem{coronavirus}
Worldometer.
\newblock Covid-19 coronavirus pandemic.
\newblock \url{https://www.worldometers.info/coronavirus/}, 2020.
\newblock Accessed: 2020-05-25.

\bibitem{zhigljavsky2020comparison}
A.~Zhigljavsky et~al.
\newblock Comparison of different exit scenarios from the lock-down for
  covid-19 epidemic in the uk and assessing uncertainty of the predictions.
\newblock {\em arXiv preprint arXiv:2004.04583}, 2020.

\bibitem{gilbert2020preparing}
M.~Gilbert et~al.
\newblock Preparing for a responsible lockdown exit strategy.
\newblock {\em Nature Medicine}, pp. 1--2, 2020.

\bibitem{funk2009spread}
S.~Funk et~al.
\newblock The spread of awareness and its impact on epidemic outbreaks.
\newblock {\em Proceedings of the National Academy of Sciences},
  106(16):6872--6877, 2009.

\bibitem{del2005effects}
S.~Del~Valle et~al.
\newblock Effects of behavioral changes in a smallpox attack model.
\newblock {\em Mathematical biosciences}, 195(2):228--251, 2005.

\bibitem{riva2014pandemic}
M.~A. Riva et~al.
\newblock Pandemic fear and literature: observations from jack london’s the
  scarlet plague.
\newblock {\em Emerging infectious diseases}, 20(10):1753, 2014.

\bibitem{poletti2011effect}
P.~Poletti et~al.
\newblock The effect of risk perception on the 2009 h1n1 pandemic influenza
  dynamics.
\newblock {\em PloS one}, 6(2), 2011.

\bibitem{perra2011towards}
N.~Perra et~al.
\newblock Towards a characterization of behavior-disease models.
\newblock {\em PloS one}, 6(8), 2011.

\bibitem{epstein2008coupled}
J.~M. Epstein et~al.
\newblock Coupled contagion dynamics of fear and disease: mathematical and
  computational explorations.
\newblock {\em PLoS One}, 3(12), 2008.

\bibitem{kim2015public}
Y.~Kim et~al.
\newblock Public risk perceptions and preventive behaviors during the 2009 h1n1
  influenza pandemic.
\newblock {\em Disaster medicine and public health preparedness},
  9(2):145--154, 2015.

\bibitem{cowling2010community}
B.~J. Cowling et~al.
\newblock Community psychological and behavioral responses through the first
  wave of the 2009 influenza a (h1n1) pandemic in hong kong.
\newblock {\em The Journal of infectious diseases}, 202(6):867--876, 2010.

\bibitem{felix2016bidirectional}
A.~C. Felix-Ortiz et~al.
\newblock Bidirectional modulation of anxiety-related and social behaviors by
  amygdala projections to the medial prefrontal cortex.
\newblock {\em Neuroscience}, 321:197--209, 2016.

\bibitem{pavlov2010conditioned}
P.~I. Pavlov.
\newblock Conditioned reflexes: an investigation of the physiological activity
  of the cerebral cortex.
\newblock {\em Annals of neurosciences}, 17(3):136, 2010.

\bibitem{maren2001neurobiology}
S.~Maren.
\newblock Neurobiology of pavlovian fear conditioning.
\newblock {\em Annual review of neuroscience}, 24(1):897--931, 2001.

\bibitem{davis1992role}
M.~Davis.
\newblock The role of the amygdala in fear and anxiety.
\newblock {\em Annual review of neuroscience}, 15(1):353--375, 1992.

\bibitem{skinner2019behavior}
B.~F. Skinner.
\newblock {\em The behavior of organisms: An experimental analysis}.
\newblock BF Skinner Foundation, 2019.

\bibitem{towers2015mass}
S.~Towers et~al.
\newblock Mass media and the contagion of fear: the case of ebola in america.
\newblock {\em PloS one}, 10(6), 2015.

\bibitem{wakefield2010use}
M.~A. Wakefield et~al.
\newblock Use of mass media campaigns to change health behaviour.
\newblock {\em The Lancet}, 376(9748):1261--1271, 2010.

\bibitem{collinson2014modelling}
S.~Collinson and J.~M. Heffernan.
\newblock Modelling the effects of media during an influenza epidemic.
\newblock {\em BMC public health}, 14(1):376, 2014.

\bibitem{collinson2015effects}
S.~Collinson et~al.
\newblock The effects of media reports on disease spread and important public
  health measurements.
\newblock {\em PloS one}, 10(11), 2015.

\bibitem{yan2016media}
Q.~Yan et~al.
\newblock Media coverage and hospital notifications: Correlation analysis and
  optimal media impact duration to manage a pandemic.
\newblock {\em Journal of theoretical biology}, 390:1--13, 2016.

\bibitem{bakshy2012role}
E.~Bakshy et~al.
\newblock The role of social networks in information diffusion.
\newblock In {\em Proceedings of the 21st international conference on World
  Wide Web}, pp. 519--528, 2012.

\bibitem{prem2017projecting}
K.~Prem et~al.
\newblock Projecting social contact matrices in 152 countries using contact
  surveys and demographic data.
\newblock {\em PLoS computational biology}, 13(9):e1005697, 2017.

\bibitem{singh2020age}
R.~Singh and R.~Adhikari.
\newblock Age-structured impact of social distancing on the covid-19 epidemic
  in india.
\newblock {\em arXiv preprint arXiv:2003.12055}, 2020.

\bibitem{kermack1927contribution}
W.~O. Kermack and A.~G. McKendrick.
\newblock A contribution to the mathematical theory of epidemics.
\newblock {\em Proceedings of the royal society of london. Series A, Containing
  papers of a mathematical and physical character}, 115(772):700--721, 1927.

\bibitem{tang2020updated}
B.~Tang et~al.
\newblock An updated estimation of the risk of transmission of the novel
  coronavirus (2019-ncov).
\newblock {\em Infectious disease modelling}, 5:248--255, 2020.

\bibitem{eikenberry2020mask}
S.~E. Eikenberry et~al.
\newblock To mask or not to mask: Modeling the potential for face mask use by
  the general public to curtail the covid-19 pandemic.
\newblock {\em Infectious Disease Modelling}, 2020.

\bibitem{schultz2006behavioral}
W.~Schultz.
\newblock Behavioral theories and the neurophysiology of reward.
\newblock {\em Annu. Rev. Psychol.}, 57:87--115, 2006.

\bibitem{humphreys1939effect}
L.~G. Humphreys.
\newblock The effect of random alternation of reinforcement on the acquisition
  and extinction of conditioned eyelid reactions.
\newblock {\em Journal of Experimental Psychology}, 25(2):141, 1939.

\bibitem{televisionmedia}
BARC.
\newblock Broadcast audience research council india.
\newblock
  \url{https://www.barcindia.co.in/resources/pdf/BARC%20India%20Universe%20Update%20-%202018.pdf},
  2018.
\newblock Accessed: 2020-05-14.

\bibitem{communityradiomedia}
AMS.
\newblock Study on listenership, reach and effectiveness of community radio
  stations in india.
\newblock
  \url{https://mib.gov.in/sites/default/files/AMS%20Report%20on%20CRS.pdf},
  2018.
\newblock Accessed: 2020-05-13.

\bibitem{internetmedia}
IAMAI.
\newblock India internet 2019.
\newblock
  \url{https://cms.iamai.in/Content/ResearchPapers/d3654bcc-002f-4fc7-ab39-e1fbeb00005d.pdf},
  2019.
\newblock Accessed: 2020-05-13.

\bibitem{newspapermedia}
MRUC.
\newblock Indian readership survey.
\newblock
  \url{https://mruc.net/uploads/posts/195a27e70e0ddae2aa3601970e191531.pdf},
  2019.
\newblock Accessed: 2020-05-12.

\bibitem{trustmedia}
IPSOS.
\newblock Trust in the media.
\newblock
  \url{https://www.ipsos.com/sites/default/files/ct/news/documents/2019-06/global-advisor-trust-in-media-2019.pdf},
  2019.
\newblock Accessed: 2020-05-13.

\bibitem{kucharski2020early}
A.~J. Kucharski et~al.
\newblock Early dynamics of transmission and control of covid-19: a
  mathematical modelling study.
\newblock {\em The lancet infectious diseases}, 2020.

\bibitem{covid19}
COVID-19.
\newblock Covid-19 india.
\newblock \url{www.covid19india.org}, 2020.
\newblock Accessed: 2020-05-14.

\bibitem{singh2015modelling}
V.~Singh.
\newblock Modelling methodologies for systems biology.
\newblock In {\em Systems and Synthetic Biology}, pp. 43--62. Springer,
  Dordrecht, 2015.

\bibitem{population}
PopulationPyramid.
\newblock Population pyramids of the world from 1950 to 2100.
\newblock \url{https://www.populationpyramid.net/india/2019/}, 2019.
\newblock Accessed: 2020-05-13.

\bibitem{cauchemez2008estimating}
S.~Cauchemez et~al.
\newblock Estimating the impact of school closure on influenza transmission
  from sentinel data.
\newblock {\em Nature}, 452(7188):750--754, 2008.

\bibitem{onder2020case}
G.~Onder et~al.
\newblock Case-fatality rate and characteristics of patients dying in relation
  to covid-19 in italy.
\newblock {\em Jama}, 2020.

\bibitem{cdc}
CDC.
\newblock Centers for disease control and prevention.
\newblock \url{https://www.cdc.gov/nchs/nvss/vsrr/covid19/index.htm}, 2020.
\newblock Accessed: 2020-05-14.

\bibitem{who}
WHO.
\newblock World health organisation.
\newblock
  \url{https://www.who.int/publications-detail/report-of-the-who-china-joint-mission-on-coronavirus-disease-2019-(covid-19)},
  2020.
\newblock Accessed: 2020-05-14.

\bibitem{dowd2020demographic}
J.~B. Dowd et~al.
\newblock Demographic science aids in understanding the spread and fatality
  rates of covid-19.
\newblock {\em Proceedings of the National Academy of Sciences},
  117(18):9696--9698, 2020.

\bibitem{obregon2009achieving}
R.~Obreg{\'o}n et~al.
\newblock Achieving polio eradication: a review of health communication
  evidence and lessons learned in india and pakistan.
\newblock {\em Bulletin of the World Health Organization}, 87:624--630, 2009.

\bibitem{stein2006specific}
D.~J. Stein and H.~Matsunaga.
\newblock Specific phobia: a disorder of fear conditioning and extinction.
\newblock {\em CNS spectrums}, 11(4):248--251, 2006.

\bibitem{milad2013deficits}
M.~R. Milad et~al.
\newblock Deficits in conditioned fear extinction in obsessive-compulsive
  disorder and neurobiological changes in the fear circuit.
\newblock {\em JAMA psychiatry}, 70(6):608--618, 2013.

\bibitem{viner2020school}
R.~M. Viner et~al.
\newblock School closure and management practices during coronavirus outbreaks
  including covid-19: a rapid systematic review.
\newblock {\em The Lancet Child \& Adolescent Health}, 2020.

\bibitem{dhar2020critique}
A.~Dhar.
\newblock A critique of the covid-19 analysis for india by singh and adhikari.
\newblock {\em arXiv preprint arXiv:2004.05373}, 2020.

\end{thebibliography}
\bibliographystyle{unsrt2authabbrvpp}

\end{document}